\documentclass[]{emulateapj}

\usepackage{natbib}
\usepackage{natbib}
\usepackage{amsmath}
\usepackage{graphicx}
\usepackage{hyperref}

\def\nicefrac#1#2{
    \raise.5ex\hbox{#1}%
    \kern-.1em/\kern-.15em%
    \lower.25ex\hbox{#2}}

\begin{document}



\title{LOFAR, VLA, and Chandra observations of the Toothbrush galaxy cluster}
\author{R.~J.~van~Weeren\altaffilmark{1,$\star$}, G.~Brunetti\altaffilmark{2},  M.~Br\"uggen\altaffilmark{3}, F.~Andrade-Santos\altaffilmark{1}, G.~A.~Ogrean\altaffilmark{1,$\dagger$}, W.~L.~Williams\altaffilmark{4,5,6}, H.~J.~A.~R\"ottgering\altaffilmark{4}, 
W.~A.~Dawson\altaffilmark{7}, W.~R.~Forman\altaffilmark{1},   F.~de~Gasperin\altaffilmark{3,4}, M.~J.~Hardcastle\altaffilmark{6}, C.~Jones\altaffilmark{1}, G.~K.~Miley\altaffilmark{4}, D.~A.~Rafferty\altaffilmark{3}, L.~Rudnick\altaffilmark{8}, J.~Sabater\altaffilmark{9}, C.~L.~Sarazin\altaffilmark{10},   T.~W.~Shimwell\altaffilmark{4},  A.~Bonafede\altaffilmark{3}, P.~N.~Best\altaffilmark{9}, L.~B{\^i}rzan\altaffilmark{3}, R.~Cassano\altaffilmark{2}, K.~T. Chy\.zy\altaffilmark{11}, J.~H.~Croston\altaffilmark{12},  T.~J.~Dijkema\altaffilmark{5},  T.~En{\ss}lin\altaffilmark{13}, C.~Ferrari\altaffilmark{14},  G.~Heald\altaffilmark{5,15}, M.~Hoeft\altaffilmark{16}, C.~Horellou\altaffilmark{17}, M.~J.~Jarvis\altaffilmark{18,19}, R.~P.~Kraft\altaffilmark{1}, M.~Mevius\altaffilmark{5},  H.~T.~Intema\altaffilmark{20,4}, S.~S.~Murray\altaffilmark{1,21},  E.~Orr\'u\altaffilmark{5}, R.~Pizzo\altaffilmark{5}, S.~S.~Sridhar\altaffilmark{15,5}, A.~Simionescu\altaffilmark{22}, A.~Stroe\altaffilmark{4}, S. van der Tol\altaffilmark{5}, and G.~J.~White\altaffilmark{23,24}\vspace{3mm}
}

\affil{\altaffilmark{1}Harvard-Smithsonian Center for Astrophysics, 60 Garden Street, Cambridge, MA 02138, USA}
\affil{\altaffilmark{2}INAF/Istituto di Radioastronomia, via Gobetti 101, I-40129 Bologna, Italy}
\affil{\altaffilmark{3}Hamburger Sternwarte, Gojenbergsweg 112, 21029 Hamburg, Germany}
\affil{\altaffilmark{4}Leiden Observatory, Leiden University, P.O. Box 9513, NL-2300 RA Leiden, The Netherlands}
\affil{\altaffilmark{5}ASTRON, the Netherlands Institute for Radio Astronomy, Postbus 2, 7990 AA, Dwingeloo, The Netherlands}
\affil{\altaffilmark{6}School of Physics, Astronomy and Mathematics, University of Hertfordshire, College Lane, Hatfield AL10 9AB, UK}
\affil{\altaffilmark{7}Lawrence Livermore National Lab, 7000 East Avenue, Livermore, CA 94550, USA}
\affil{\altaffilmark{8}Minnesota Institute for Astrophysics, University of Minnesota, 116 Church St. S.E., Minneapolis, MN 55455, USA}
\affil{\altaffilmark{9}Institute for Astronomy, University of Edinburgh, Royal Observatory, Blackford Hill, Edinburgh EH9 3HJ, UK}
\affil{\altaffilmark{10}Department of Astronomy, University of Virginia, Charlottesville, VA 22904-4325, USA}
\affil{\altaffilmark{11}Astronomical Observatory, Jagiellonian University, ul. Orla 171, 30-244 Krak\'ow, Poland}
\affil{\altaffilmark{12}School of Physics and Astronomy, University of Southampton, Southampton SO17 1BJ, UK}
\affil{\altaffilmark{13}Max Planck Institute for Astrophysics, Karl-Schwarzschildstr. 1, 85741 Garching, Germany}
\affil{\altaffilmark{14}Laboratoire Lagrange, Universit\'e C\^{o}te d'Azur, Observatoire de la C\^{o}te d'Azur, CNRS, Blvd de l'Observatoire, CS 34229, 06304 Nice cedex 4, France}
\affil{\altaffilmark{15}Kapteyn Astronomical Institute, P.O. Box 800, 9700 AV Groningen, The Netherlands}
\affil{\altaffilmark{16}Th\"uringer Landessternwarte Tautenburg, Sternwarte 5, 07778, Tautenburg, Germany}
\affil{\altaffilmark{17}Department of Earth and Space Sciences, Chalmers University of Technology, Onsala Space Observatory, SE-43992 Onsala, Sweden} 
\affil{\altaffilmark{18}Oxford Astrophysics, Department of Physics, Keble Road, Oxford, OX1 3RH, UK}
\affil{\altaffilmark{19}University of the Western Cape, Bellville 7535, South Africa}
\affil{\altaffilmark{20}National Radio Astronomy Observatory, 1003 Lopezville Road, Socorro, NM 87801-0387, USA}
\affil{\altaffilmark{21}Department of Physics and Astronomy, Johns Hopkins University, 3400 North Charles Street, Baltimore, MD 21218, USA}
\affil{\altaffilmark{22}Institute of Space and Astronautical Science (ISAS), JAXA, 3-1-1 Yoshinodai, Chuo, Sagamihara, Kanagawa, 252-5210, Japan}
\affil{\altaffilmark{23}Department of Physical Sciences, The Open University, Walton Hall, Milton Keynes MK7 6AA, UK}
\affil{\altaffilmark{24}RALSpace, The Rutherford Appleton Laboratory, Chilton, Didcot, Oxfordshire OX11 0NL, UK}

\email{E-mail: rvanweeren@cfa.harvard.edu}

\altaffiltext{$\star$}{Einstein Fellow}
\altaffiltext{$\dagger$}{Hubble Fellow}

\shorttitle{LOFAR, VLA, and Chandra observations of the Toothbrush}
\shortauthors{van Weeren et al.}

\vspace{0.5cm}
\begin{abstract}
\noindent

We present deep LOFAR observations between 120 --181 MHz of the `Toothbrush' (RX J0603.3+4214), a cluster that contains one of the brightest  radio relic sources known.  Our LOFAR observations exploit a new and novel calibration scheme to probe 10 times deeper than any previous study in this relatively  unexplored  part of the spectrum. The LOFAR observations, when combined with  VLA, GMRT, and {\it Chandra} X-ray data, provide new information about the nature of cluster merger shocks and their role in re-accelerating relativistic particles.   We derive a spectral index of $\alpha = -0.8 \pm 0.1$ at the northern edge of the main radio relic, steepening towards the south to $\alpha \approx - 2$. The spectral index of the radio halo is remarkably uniform ($\alpha = -1.16$, with an intrinsic scatter of $\leq 0.04$). The observed radio relic spectral index gives a Mach number of $\mathcal{M} = 2.8^{+0.5}_{-0.3}$, assuming diffusive shock acceleration (DSA).  However, the gas density jump at the northern edge of the large radio relic implies a much weaker shock ($\mathcal{M} \approx 1.2$, with an upper limit of $\mathcal{M} \approx 1.5$). The discrepancy between the Mach numbers calculated from the radio and X-rays can be explained if either (i) the relic traces a complex shock surface along the line of sight, or (ii) if the radio relic emission is produced by a re-accelerated population of fossil particles from a radio galaxy. Our results highlight the need for additional theoretical work and  numerical  simulations of particle acceleration and re-acceleration at cluster merger shocks.

\vspace{4mm}
\end{abstract}



\section{Introduction}
Merging  clusters of galaxies are excellent laboratories to study cosmic ray (CR) acceleration in dilute cosmic plasmas, investigate the self-interaction properties of dark matter, and probe the physics of shocks and cold fronts in the intracluster medium (ICM). Diffuse radio emission has been observed in the majority of massive merging clusters, indicating the presence of relativistic particles (i.e., cosmic rays) and magnetic fields. Given the large extent  of these sources ($\sim 1$~Mpc) and the limited lifetime of the synchrotron emitting electrons, these relativistic particles must be created or (re)-accelerated in-situ \citep[e.g.,][]{1977ApJ...212....1J}.

The diffuse cluster radio sources in merging clusters are referred to as radio halos and relics, depending on their location, morphology, and polarization properties. During the last decade, considerable progress has been made in our understanding of these sources \citep[for reviews see][]{2012A&ARv..20...54F,2014IJMPD..2330007B}.

Radio halos are centrally located in merging galaxy clusters and have typical sizes of about one Mpc. They have smoother brightness distributions than radio relics and the radio emission roughly follows the X-ray emission from the thermal ICM. Halos are unpolarized above the few percent level, although a few cases of polarized emission from halos have been reported. However, the nature of the polarized emission is still being debated \citep[e.g.,][]{2011A&A...525A.104P,2005A&A...430L...5G}.

For radio halos,  a correlation has been found between cluster X-ray luminosity ($L_{\rm{X}}$) and radio halo power  \citep[e.g.,][]{2013ApJ...777..141C,2010A&A...517A..10C}. A similar correlation with radio power is found  \citep[e.g.,][]{2012MNRAS.421L.112B,2013ApJ...777..141C} for the integrated Sunyaev-Zel'dovich (SZ) effect signal, given by $Y_{\rm{SZ}}$. $Y_{\rm{SZ}}$ traces the integrated pressure along the line of sight. Both $L_{\rm{X}}$ and $Y_{\rm{SZ}}$ are used as observational proxies of cluster mass. 

Two main physical models to explain the presence of radio halos have been proposed. In the secondary (or hadronic) model, the CR electrons are secondary products of collisions between thermal ions and relativistic protons in the ICM \citep[e.g.,][]{1980ApJ...239L..93D, 1999APh....12..169B, 2000A&A...362..151D, 2001ApJ...562..233M,2004A&A...413...17P,2008MNRAS.385.1211P,2010ApJ...722..737K,2011A&A...527A..99E,2014MNRAS.438..124Z}. In the turbulent re-acceleration model, electrons are re-accelerated by merger induced magnetohydrodynamical turbulence \citep[e.g.,][]{1987A&A...182...21S,2001MNRAS.320..365B, 2001ApJ...557..560P,2015arXiv150307870P}. Secondary models are challenged by the large energy content of cosmic ray protons needed to explain radio halos with very steep spectra \citep[][]{2004JKAS...37..493B, 2008Natur.455..944B} and by the non-detection of $\gamma$-rays \citep[e.g.,][]{2011ApJ...728...53J,2012MNRAS.426..956B,2014ApJ...787...18A}.

Radio relics are elongated and often arc-like sources found in the outskirts of galaxy clusters.
There is a considerable amount of evidence that radio relics trace shock fronts where particles are being  (re-)accelerated, as was proposed
initially in \cite{1998A&A...332..395E}. The key observational facts that support the shock-radio relic connection are (i) the high polarization of some relics, with the apparent magnetic field lines being parallel to the major axis of the relics \citep[e.g.,][]{2010Sci...330..347V,2012MNRAS.426...40B,2012MNRAS.426.1204K,2014MNRAS.444.3130D}, indicating that the ICM and associated magnetic fields are compressed, (ii) the presence of spectral index gradients, indicating electron cooling in the post-shock region of an outward traveling shock wave \citep[e.g.,][]{2008A&A...486..347G,2010Sci...330..347V,2011A&A...528A..38V,2013A&A...555A.110S,2014MNRAS.445..330H}, (iii) a change from power-law radio spectra at the outer edge of relics towards curved spectra in the post-shock region, indicating a site of acceleration and electron cooling \citep{2012A&A...546A.124V,2013A&A...555A.110S}, and  (iv) the presence of ICM density and/or temperature jumps at the location of relics \citep[e.g.,][]{2010ApJ...715.1143F,2011ApJ...728...82M,2013ApJ...764...82B,2013MNRAS.433.1701O,2013PASJ...65...16A,2014MNRAS.440.3416O}. X-ray observations indicate that the shock Mach numbers ($\mathcal{M}$) are low, typically  $\lesssim 3$.

Although the above observations provide support for the basic scenario where particles are accelerated by outward traveling shock waves, there are still several open questions and observational puzzles. One of these puzzles concerns the low Mach numbers. The efficiency of acceleration via the diffusive shock acceleration (DSA) mechanism \citep[e.g.,][]{1983RPPh...46..973D} at these weak shocks in clusters is thought to be very low. A possible solution to this problem is that the shock re-accelerates a population of already mildly relativistic electrons, instead of directly accelerating them from the thermal pool \citep[e.g.,][]{2005ApJ...627..733M,2011ApJ...734...18K,2012ApJ...756...97K}. There is some observational support for this shock re-acceleration model \cite[][]{2013ApJ...769..101V,2014ApJ...785....1B,2015MNRAS.449.1486S} and this would help to explain the existence of bright radio relics \citep{2013MNRAS.435.1061P}. 

Particle-in-cell (PIC) simulations can provide crucial information  about electron and proton acceleration at shocks. Recent PIC simulations  indicate that instead of DSA,  shock drift acceleration (SDA) is a more important mechanism at the initial stages of the acceleration process \citep{2014ApJ...794..153G,2014ApJ...783...91C,2014ApJ...797...47G}. 
In these simulations, SDA results in efficient acceleration by low-Mach number shocks in typical ICM conditions and thus potentially solves the low-acceleration efficiency problem of DSA.

Another observational puzzle is that some clusters host shocks that do not show corresponding radio relics \citep{2014MNRAS.440.2901S,2011MNRAS.417L...1R}. The opposite situation, i.e., relics without clear shocks, is also observed \citep[e.g.,][]{2014MNRAS.443.2463O}. However, it is important to note that the X-ray detection of shocks in cluster outskirts, in particular those with low-Mach numbers ($\mathcal{M} \lesssim 2$), is difficult. Given that radio relics seem to trace low-Mach number shocks, it is to be expected that in some cases these shocks escape detection in the X-rays, simply because of observational limitations. Some shocks coincide with the ``edges'' of radio halos \citep[e.g.,][]{2005ApJ...627..733M,2014MNRAS.440.2901S}, suggesting a relation between shocks and halos.

 
 \subsection{The `Toothbrush' galaxy cluster}
 The `Toothbrush' galaxy cluster (RX~J0603.3+4214) hosts one of the largest and brightest radio relics, and therefore serves as an excellent target for a low-frequency radio study. The cluster was discovered by \cite{2012A&A...546A.124V} and is located at $z=0.225$ \citep{dawson}.  Besides the bright toothbrush-shaped radio relic, the cluster hosts at least two additional fainter relics and an elongated radio halo. The integrated Toothbrush radio relic spectrum has a power-law shape\footnote{$F_{\nu} \propto \nu^{\alpha}$, where $\alpha$ is the spectral index} between 74~MHz and 4.9~GHz with $\alpha = -1.10 \pm 0.02$. A clear north-south spectral index gradient is also observed for the relic. In addition, the radio spectra at the relic's outer edge are straight and relatively flat, while the spectrum curves in regions with a steeper spectral index. From a radio color-color analysis \citep{1993ApJ...407..549K} an injection spectral index of $-0.65\pm0.05$ was found, corresponding to a Mach number of  3.3--4.6 (under the assumption of DSA in the linear test particle regime). The radio observations thus suggest that the Toothbrush traces a moderately strong, northwards-moving shock, that accelerates particles which later lose energy in the post-shock region.
 
 Numerical simulations suggest that a triple merger event causes the peculiar shape of the putative shock that produces the Toothbrush relic \citep{2012MNRAS.425L..76B}.
 This merger scenario consists of a main north-south merger event, together with a third less massive substructure moving in from the southwest.  Although {\it XMM-Newton} observations  revealed that  the main merger event takes place along a north-south axis, no evidence was found for a third smaller substructure \citep{2013MNRAS.433..812O}. 
Several candidate shocks were identified in the cluster from X-ray surface brightness edges. In the north, a weak edge was found, but the location of the edge was offset from the expected shock location based on the relic position. \cite{dawson}  find additional substructure in the cluster which indicates that the merger is likely more complex than a simple north-south event.

Deep H$\alpha$  imaging of the {Toothbrush cluster and CIZA~J2242.8+5301} was presented by \cite{2014MNRAS.438.1377S}.  No boost in the number of  H$\alpha$ emitters was found in the Toothbrush cluster, in stark contrast with the merging cluster CIZA~J2242.8+5301.   

Here we present  LOw Frequency ARray (LOFAR) observations of the Toothbrush galaxy cluster. LOFAR is a powerful new radio telescope operating between 10 to 240~MHz \citep{2013A&A...556A...2V} which allows studies of diffuse cluster radio emission with high-angular resolution and sensitivity. {To produce high-quality LOFAR images we developed a new calibration scheme which is presented in a separate paper. This calibration scheme will serve as a reference for future deep observations and surveys carried out with LOFAR.} We combine the LOFAR data with Karl G. Jansky Very Large Array (VLA) {1--2}~GHz L-band observations and previously published Giant Metrewave Radio Telescope (GMRT) data to study the radio spectral index.
We also present new deep {\it Chandra} ACIS-I observations of the cluster. This allows us to search for surface brightness edges associated with shocks and cold fronts, and to map the ICM temperature distribution. 
 
The layout of this paper is as follows. {We start with an overview of the observations and data reduction in Section~\ref{sec:obs}.} We present the radio and X-ray images, together with the spectral index and temperatures maps in Sect.~\ref{sec:results}. The results of our search for shocks and cold fronts are described in Sect.~\ref{sec:edges}. This is followed by a discussion and conclusions in Sects.~\ref{sec:dicussions} and \ref{sec:conclusions}. 

In this paper, we adopt a $\Lambda$CDM cosmology with $H_{0} = 70$~km~s$^{-1}$~Mpc$^{-1}$, $\Omega_{m} = 0.3$, and $\Omega_{\Lambda} = 0.7$. With the adopted cosmology, 1\arcsec~corresponds to a physical scale of 3.614~kpc at $z=0.225$, the redshift of the Toothbrush cluster.

\section{Observations}
\label{sec:obs}

\subsection{LOFAR HBA Observations}

The Toothbrush cluster was observed on Feb 24, 2013, for 10~hrs with the LOFAR High Band Antenna (HBA) stations. 
Complete frequency coverage between 112--181 MHz was obtained on the Toothbrush field. The nearby calibrator 3C147 was simultaneously observed using a second station beam.

For the observations {13 remote and 42 (split) core} stations were used \citep[see][for a description of the stations]{2013A&A...556A...2V}, giving baselines that range between 68~m and 80~km. The large number of short baselines  is important to image diffuse extended emission.

{Advanced calibration and processing techniques are needed to obtain deep high-fidelity images at low radio frequencies. We employ the facet calibration scheme to correct  for direction dependent effects (DDEs) which are caused by the ionosphere and imperfect knowledge of the station beam shapes. Below we briefly summarize the calibration of the LOFAR data. For more details about the calibration and observations the reader is referred to \cite{vanweerenfacet}.}

{The first part of the data reduction consists of the removal of radio frequency interference (RFI), subtraction of bright off-axis sources  appearing through the station beam sidelobes, and transfer of the flux-scale and clock corrections from the calibrator source 3C147 to the target field observation. After applying these corrections, we calibrate these data against a target field model derived from GMRT 150~MHz observations \citep{2012A&A...546A.124V}. We decided to discard the data below 120~MHz since these data are generally more noisy and harder to calibrate. In the second part of the calibration, we obtain DDE corrections towards about 70~directions, so-called ``facets'', and apply these corrections during the imaging. With this scheme we can overcome the limits of traditional self-calibration which cannot correct for DDEs. This facet calibration scheme allows us to obtain deep ($\sim 0.1$~mJy~beam$^{-1}$) and high-resolution ($\sim 5\arcsec$) images of the cluster in the HBA band.}

\subsection{VLA observations}
\label{sec:jvladata}
VLA L-band observations were taken in D-array on Jan 28, 2013. The data were recorded with 16 spectral windows each having 64 channels, covering the 1--2~GHz band. An overview of the observations is given in Table~\ref{tab:jvlaobs}. 

\begin{table}[]
\begin{center}
\caption{VLA Observations}
\begin{tabular}{lllll}
\hline
\hline
Array configuration 			& D			& \\
Observation date 			&   28 Jan, 2013 &   \\
Frequency coverage (GHz)     	& 1--2 &  \\
On source time   (hr)   &   3.2 & \\
Correlations   & full stokes \\
Channel width (MHz) &1  \\
Integration time (s) &  5 \\
Resolution (arcsec)  & $31 \times 24$\ \\
Image noise ($\mu$Jy~beam$^{-1}$)  & 35\ \\
\hline
\hline
\end{tabular}
\label{tab:jvlaobs}
\end{center}
\end{table}

The data were reduced and calibrated with {\tt CASA} \citep{2007ASPC..376..127M}. As a first step the data were  Hanning smoothed. We removed RFI using the {\tt AOFlagger} and applied the elevation dependent gain tables and antenna offsets positions. We calibrated the antenna delays, bandpass, and polarization leakage terms using the primary calibrator 3C147. The cross-hand delays and polarization angles were set using 3C138. Gain calibration was obtained for these calibrator sources as well. The gain solutions were then transferred to the target sources. 

To refine the gain calibration several rounds of self-calibration were carried out. All imaging was done with W-projection  \citep{2008ISTSP...2..647C,2005ASPC..347...86C}, MS-MFS \citep{2011A&A...532A..71R}, and multi-scale clean \citep{2008ISTSP...2..793C}. We used three Taylor terms ({\tt nterms=3}) to take the frequency dependence of the sky into account. For the self-calibration we used \cite{briggs_phd} weighting, with a {\tt robust} factor of 0.0. Clean masks were also employed. The final images were corrected for the primary beam attenuation.

\subsection{Chandra observations}
\label{sec:chandrareduction}
 \textit{Chandra} ACIS-I observations totaling 237~ks were taken in late 2013 (\dataset [ADS/Sa.CXO#obs/15171,ADS/Sa.CXO#obs/15172,ADS/Sa.CXO#obs/15323] {ObsID: 15171, 15172, 15323}). The data were  calibrated with the {\tt chav} package\footnote{http://hea-www.harvard.edu/\~{}alexey/CHAV/}, following the processing described in \cite{2005ApJ...628..655V}, applying the most recent calibration files\footnote{We used CIAO v4.6 and CALDB v4.6.5}. The calibration and data reduction consists of filtering of counts with a recomputed ASCA grades 1, 5, or~7 and those from bad pixels, a correction for position-dependent charge transfer inefficiency, and the application of gain maps to calibrate photon energies. Periods of high background were also filtered by examining the count rate in the  6--12~keV  band.  {The final exposure time available for the Chandra data after filtering was 236~ks}. Standard blank sky background files were used for background subtraction. We used a pixel binning factor of 2. For more details about the data reduction the reader is referred to \cite{2005ApJ...628..655V}.

\section{Results}
\label{sec:results}

\begin{figure*}[ht!]
\begin{center}
\includegraphics[angle =180, trim =0cm 0cm 0cm 0cm,width=0.47\textwidth]{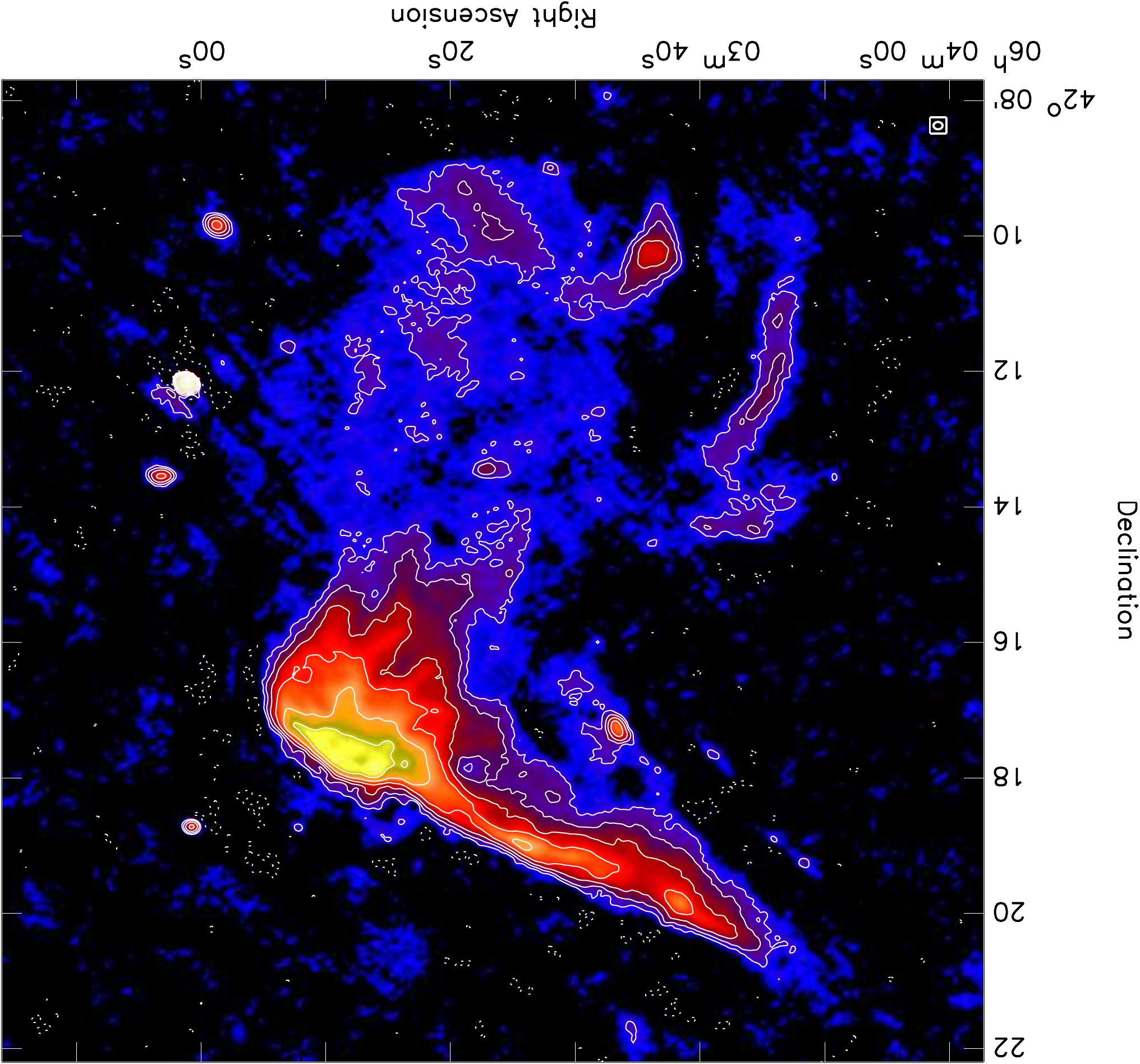}
\includegraphics[angle =180, trim =0cm 0cm 0cm 0cm,width=0.47\textwidth]{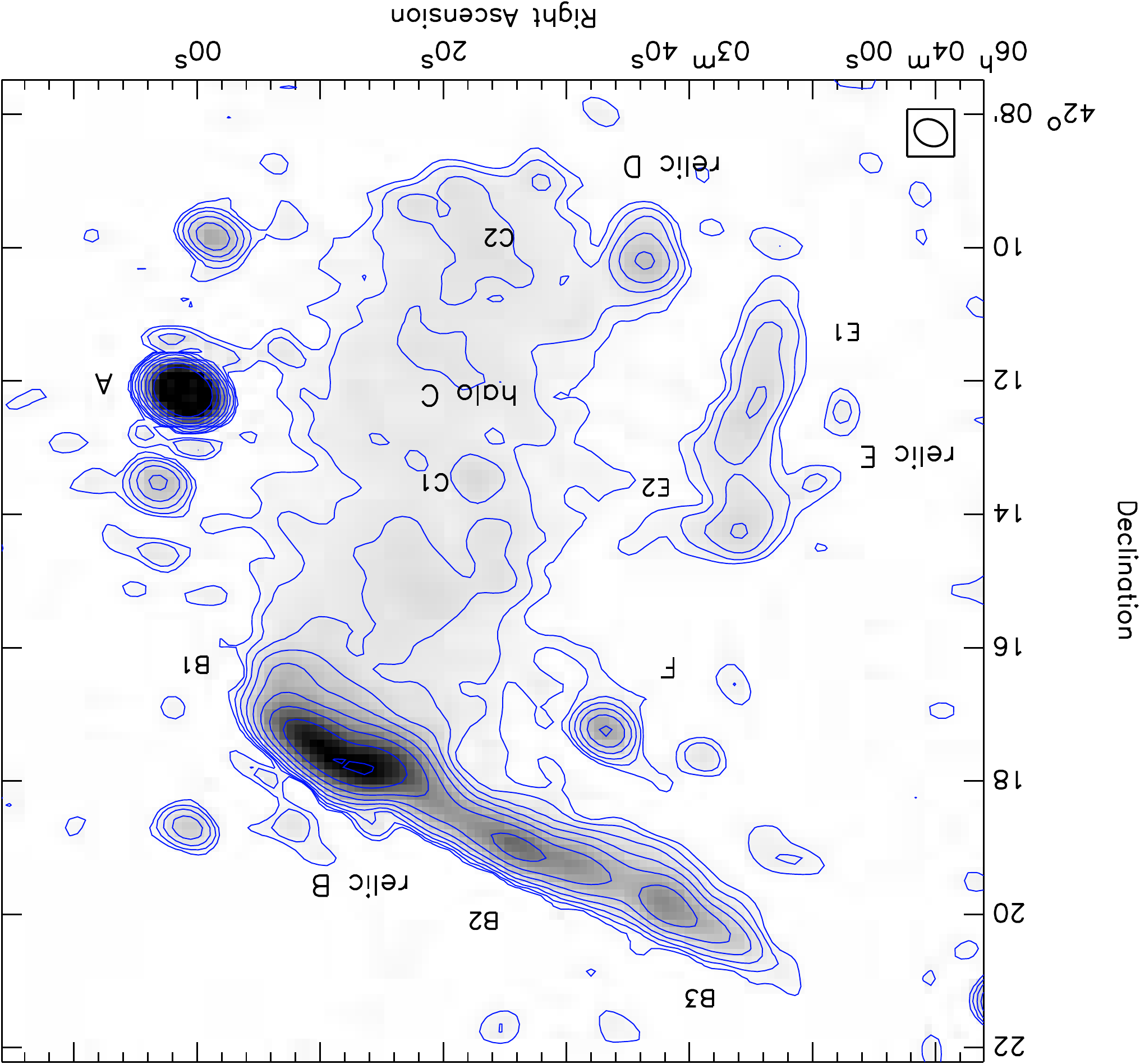}
\end{center}
\caption{Left: LOFAR HBA 120--181~MHz image. The beam size is $8.0\arcsec \times 6.5\arcsec$ as indicated in the bottom left corner. Contour levels are drawn at ${[1, 2, 4, 8, \ldots]} \times 4\sigma_{\mathrm{rms}}$, with  $\sigma_{\mathrm{rms}}=93$~$\mu$Jy~beam$^{-1}$. Negative $-3\sigma_{\mathrm{rms}}$ contours are shown with dotted lines. Right: D-array VLA  {1--2}~GHz image with various sources labelled, following \cite{2012A&A...546A.124V}. Some of the sources are further subdivided into numbered components, e.g., relic E is divided into E1 and E2. The beam size is $31\arcsec \times 24\arcsec$. Contour levels are drawn at the same levels as in the left panel, but with $\sigma_{\mathrm{rms}}=35$~$\mu$Jy~beam$^{-1}$.  }
\label{fig:hbaimage}
\end{figure*}

\subsection{LOFAR 150~MHz radio continuum images}

\begin{table}
\begin{center}
\caption{Properties of diffuse radio sources in RX~J0603.3+4214}
\tabcolsep=0.1cm
\begin{tabular}{llllll}
\hline
\hline
 & $S_{\rm{1.5~GHz}}^{a}$  &$S_{\rm{150~MHz}}^{a}$ & $P_{\rm{1.4 GHz}}$ &LLS$^{b}$ & $\alpha^{1500}_{150}$\\
              &      mJy                                   &  Jy  & $10^{24}$~W~Hz$^{-1}$  &Mpc    \\
\hline
B&$317\pm31$ & $3.9\pm0.4$ &65 & 1.9 & $-1.09 \pm 0.06$ \\
C&$46\pm 5$ & $0.55\pm0.06$ &9.1& $\sim1.5$  & $-1.08 \pm 0.06$\\
D&  $6.5\pm 0.7$& $0.086\pm 0.009$ & 1.4   & 0.25  & $-1.11 \pm 0.06$\\
E& $14.3\pm0.2$ &$0.15\pm0.02$ & 2.5 & 1.1 & $-1.03 \pm 0.06$\\
\hline
\hline
\end{tabular}
\label{tab:rx42diffuse}
\end{center}
$^{a}$ Integrated flux densities were measured from the LOFAR and VLA images that were used to make Figure~\ref{fig:spixhbajvla}. The regions where the fluxes were extracted are indicated in Figure~\ref{fig:intfluxes}.\\
$^{b}$ largest linear size\\
\end{table}

Our deep $8.0\arcsec\times 6.5\arcsec$ resolution LOFAR image of the cluster, with {\tt robust$=-0.25$} weighting, is shown in Figure~\ref{fig:hbaimage} (left panel). The VLA 1--2~GHz D-array image, with a resolution of $31\arcsec\times 24\arcsec$, is shown in the right panel of Figure~\ref{fig:hbaimage}. To facilitate the discussion, we labeled the diffuse and compact sources on the VLA image, following the scheme from \cite{2012A&A...546A.124V}. The LOFAR image (Figure~\ref{fig:hbaimage}) reveals both the previously known radio relics (sources B, D, E) and the extended radio halo (source C). The precise boundary, between where the relic emission from B1 ends and the emission from radio halo C starts, is difficult to determine. {We define the boundary of the halo and relic to be located at the point where the spectral index approaches a value of about $-1.1$, the average spectral index of the radio halo, see Sect.~\ref{sec:spix} and Figure~\ref{fig:intfluxes}.} An overview of the properties of the diffuse radio sources in the cluster is given in Table~\ref{tab:rx42diffuse}.

In terms of depth and resolution {(assuming a spectral index of about $-1$)}, the LOFAR image is comparable to our deep GMRT 610~MHz image \citep{2012A&A...546A.124V}. A difference with respect to the GMRT~610~MHz observations is that the radio halo is clearly detected in the LOFAR image, while it is missing in the GMRT image. This difference is caused by the much denser inner uv-coverage of LOFAR, allowing excellent mapping of extended low-surface brightness emission. Another difference with respect to the GMRT 610~MHz observations is that  relic B is wider (i.e, extends further to the south, in particular component B1). This is expected given the steep north-south spectral index gradient.  For a comparison with our GMRT 150~MHz image from \cite{2012A&A...546A.124V}, see Figure~\ref{fig:gmrthbacomp}.

\begin{figure*}[h!]
\begin{center}
\includegraphics[angle =180, trim =0cm 0cm 0cm 0cm,width=0.95\textwidth]{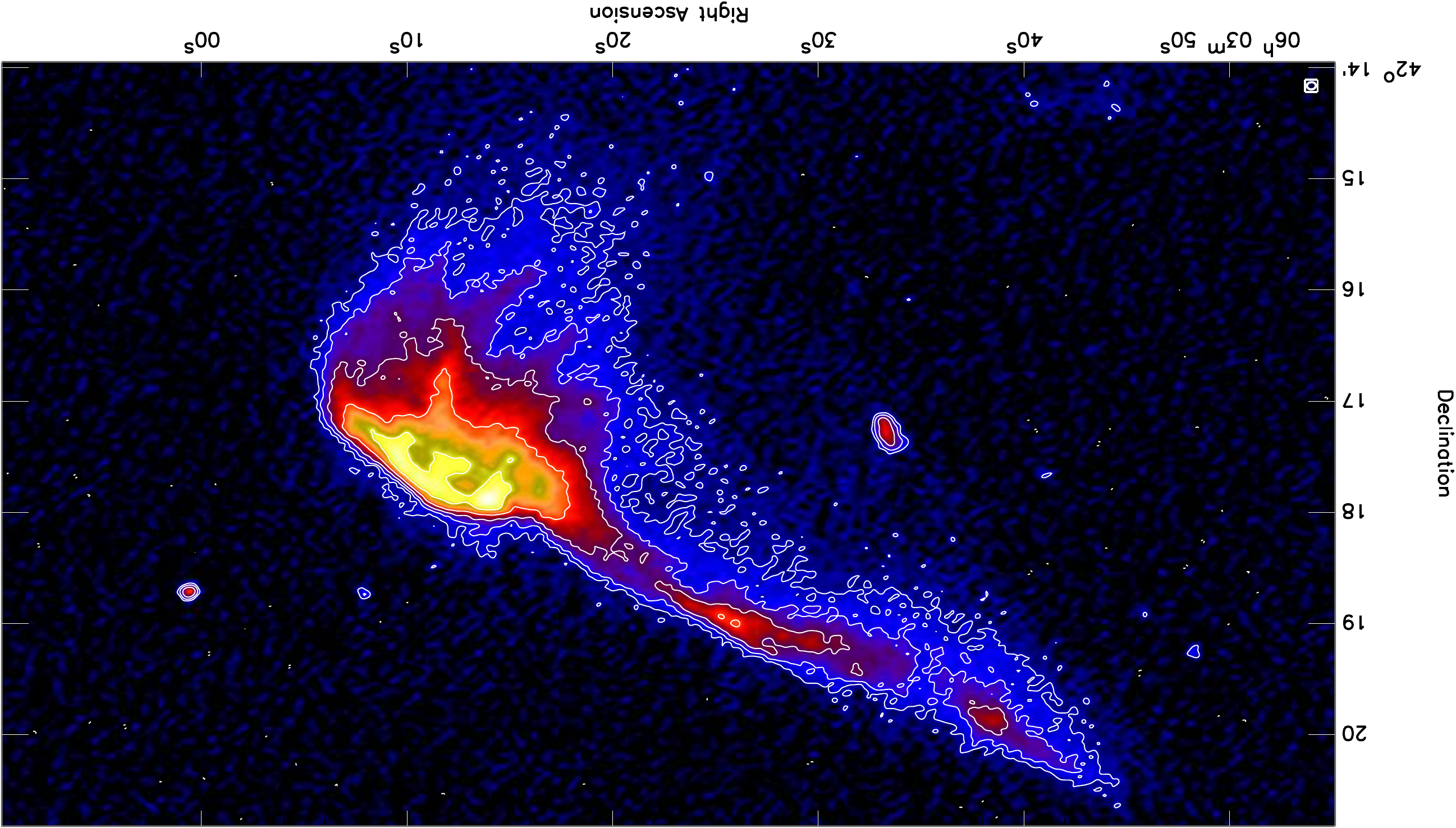}
\end{center}
\caption{LOFAR HBA 120--181~MHz images made with  {\tt robust=-1.5} (i.e., close to uniform) weighting of the visibilities. The  beam size is $4.8\arcsec\times3.9\arcsec$~and  indicated in the bottom left corner. Contour levels are drawn at ${[1, 2, 4, 8, \ldots]} \times 4\sigma_{\mathrm{rms}}$, with  $\sigma_{\mathrm{rms}}=121$~$\mu$Jy~beam$^{-1}$.  
}
\label{fig:hbaimagehighres}
\end{figure*}

We made an image with close to uniform weighting (Figure~\ref{fig:hbaimagehighres}) to better enhance small-scale details in relic B.  Several filamentary ``streams'' extend from the north of B1 to the south. These ``streams'' were also seen in the GMRT 610~MHz image and seem to ``feed'' the radio halo (C1) below B1. The northern boundary of  relic B remains relatively sharp at  4--5\arcsec~resolution. 

\begin{figure*}[hb!]
\begin{center}
\includegraphics[angle =180, trim =0cm 0cm 0cm 0cm,width=0.47\textwidth]{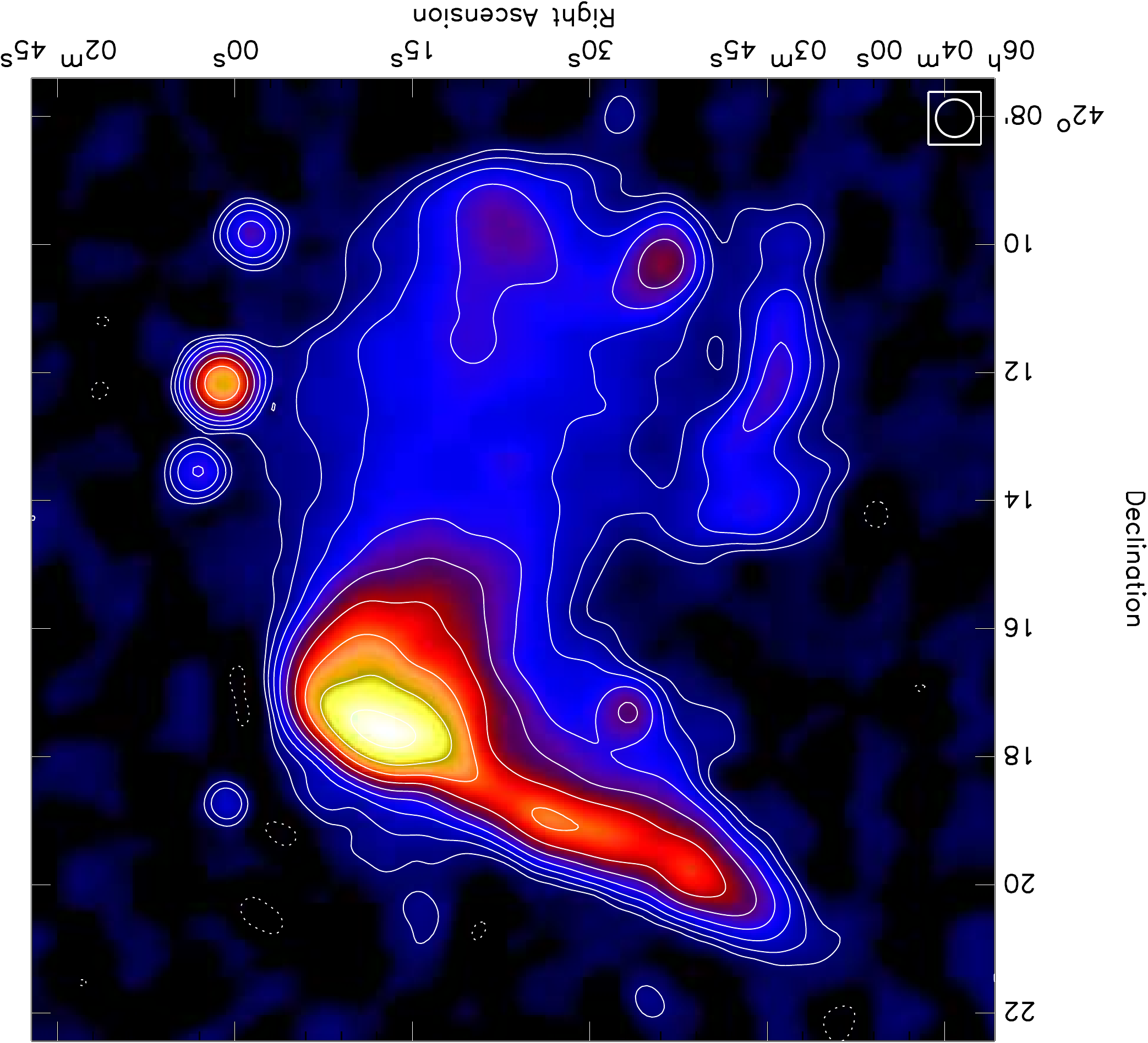}
\includegraphics[angle =180, trim =0cm 0cm 0cm 0cm,width=0.47\textwidth]{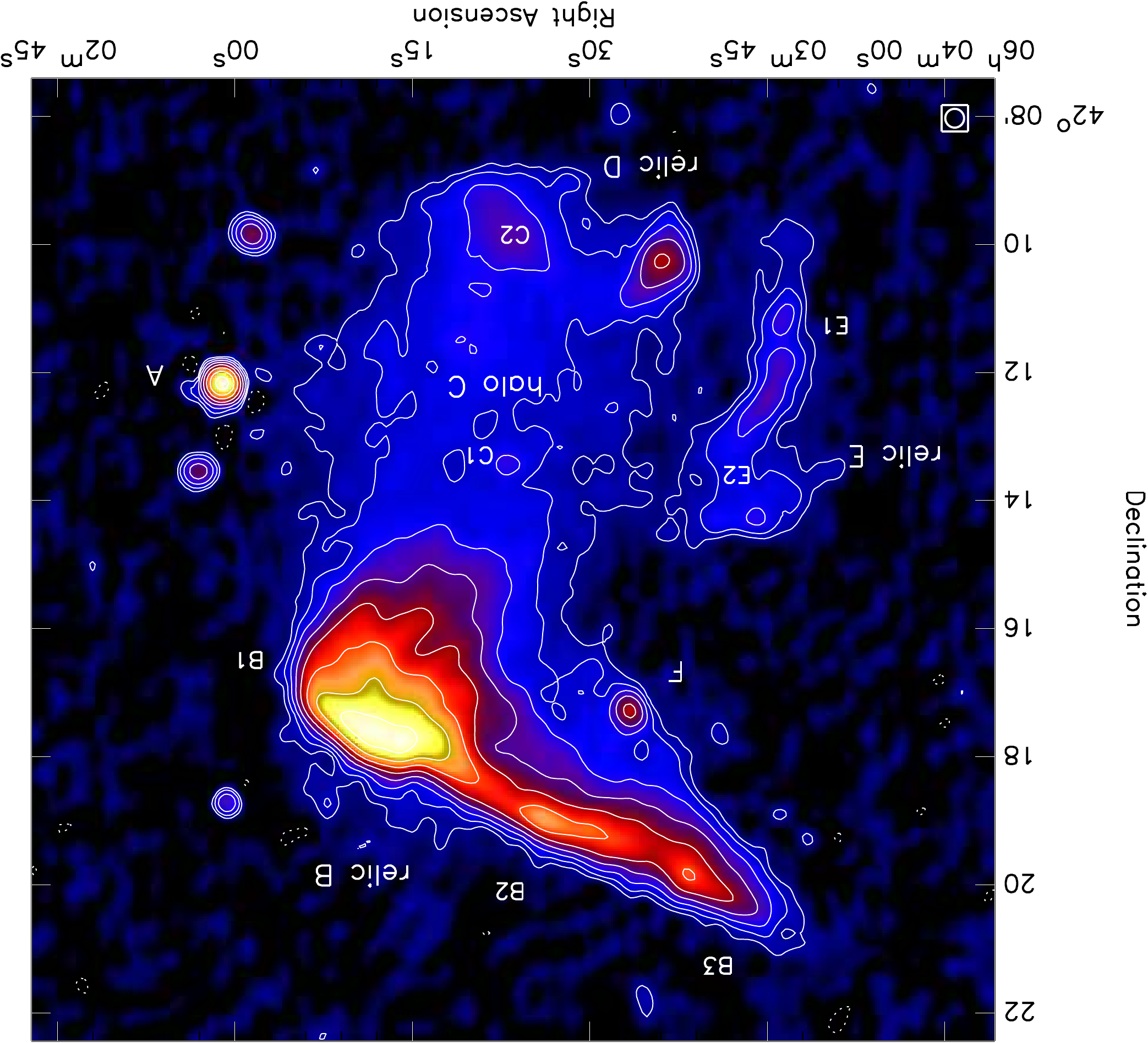}
\end{center}
\caption{LOFAR HBA 120--181~MHz images with uniform weighting and uv-tapers of 40\arcsec~(left panel) and 20\arcsec~(right panel; sources are labelled as in Figure~\ref{fig:hbaimage}). The resulting beams are close to circular with sizes of  35.2\arcsec~and 17.7\arcsec~as indicated in the bottom left corner. Contour levels are drawn at ${[1, 2, 4, 8, \ldots]} \times 4\sigma_{\mathrm{rms}}$, with  $\sigma_{\mathrm{rms}}=358$ and $212$~$\mu$Jy~beam$^{-1}$. Negative $-3\sigma_{\mathrm{rms}}$ contours are shown with dotted lines. }
\label{fig:hbaimagelowres}
\end{figure*}

\begin{figure*}[h!]
\begin{center}
\includegraphics[angle =0, trim =0cm 0cm 0cm 0cm,width=1.0\textwidth]{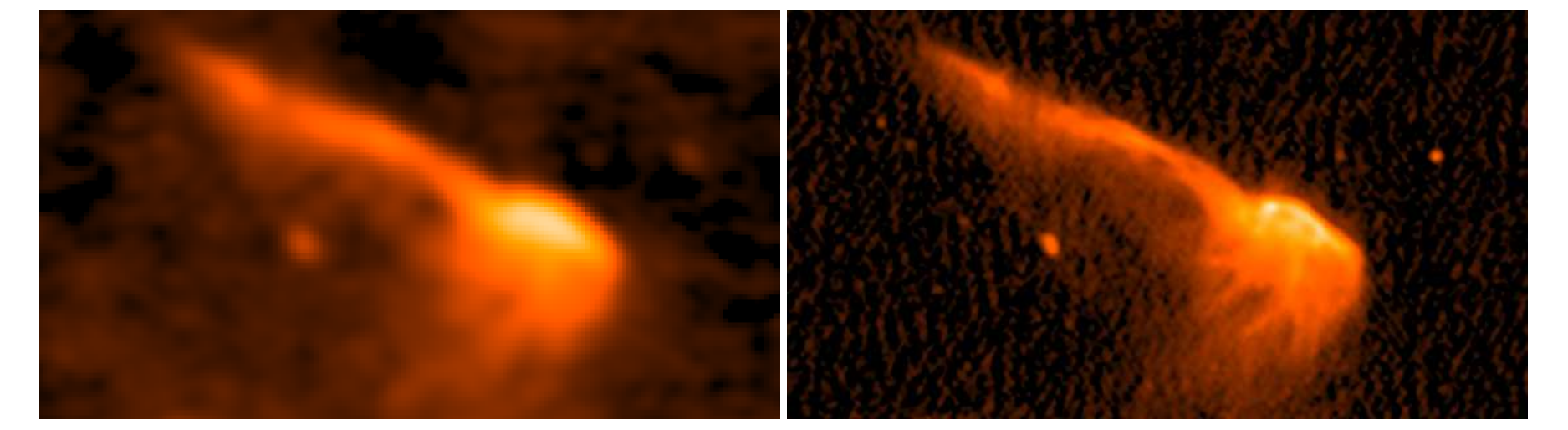} 
\end{center}
\caption{A comparison between the GMRT 150~MHz image \citep[from][]{2012A&A...546A.124V}, with a resolution of  $26\times22\arcsec$, and the 120--181~MHz LOFAR image (from Figure~\ref{fig:hbaimagehighres}), with a resolution of $4.8\arcsec\times3.9\arcsec$. The colors in both images were scaled manually to achieve similar image surface brightness levels. Except for the fine-scale structures, the images show very good agreement.}
\label{fig:gmrthbacomp}
\end{figure*}

To better bring out the radio halo, we re-imaged the dataset with uniform weighting and tapered with Gaussians in the uv-plane corresponding to resolutions of 20\arcsec~and 40\arcsec. These two images are shown in Figure~\ref{fig:hbaimagelowres}. At these resolutions, relic B extends as far south as source F and faint emission connects the halo with relic E. The southern part of the halo, C2, has an enhanced surface brightness compared to the northern part C1.

\subsection{Spectral indices}
\label{sec:spix}

\begin{figure*}[t]
\begin{center}
\includegraphics[angle =180, trim =0cm 0cm 0cm 0cm,width=0.49\textwidth]{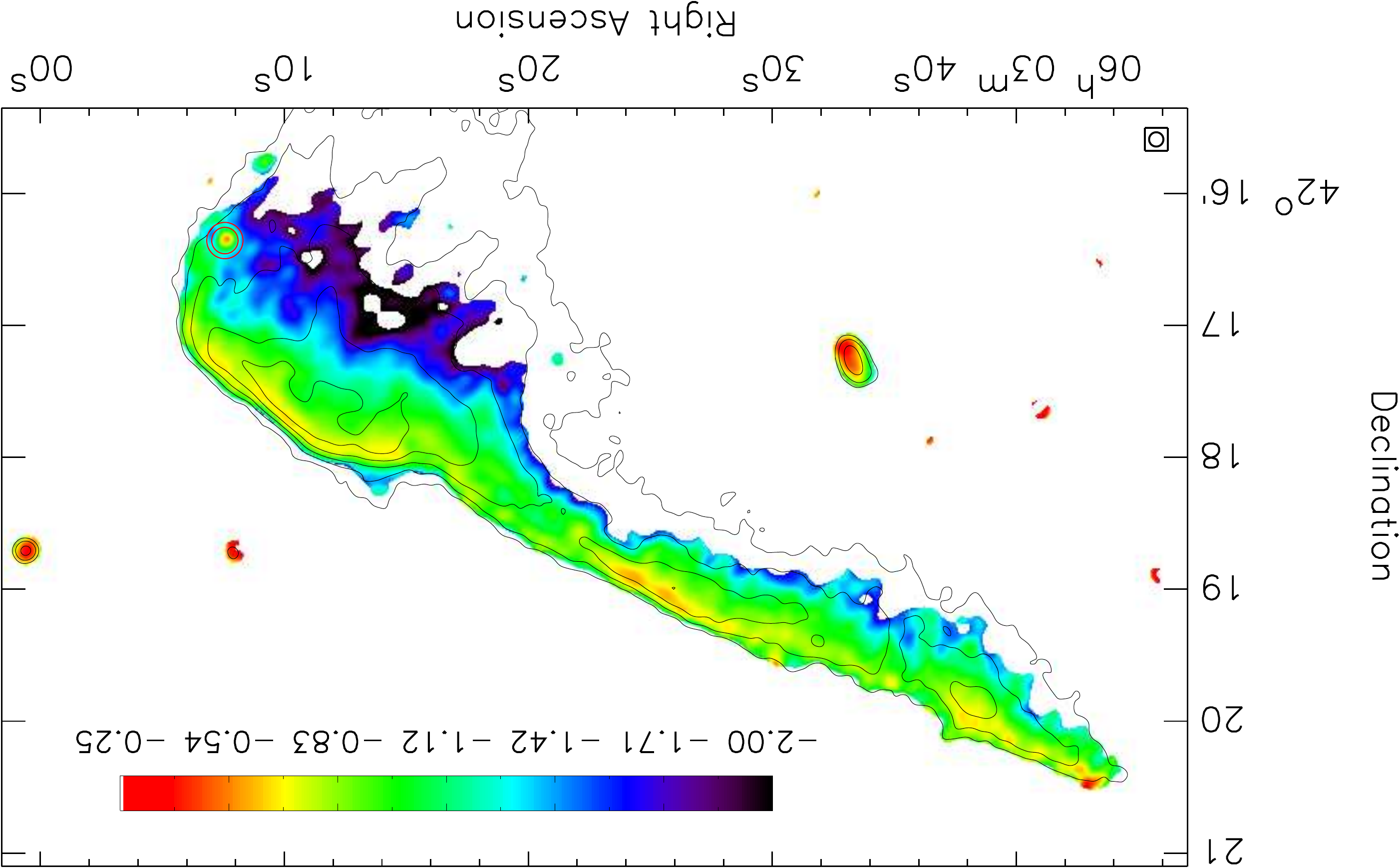}
\includegraphics[angle =180, trim =0cm 0cm 0cm 0cm,width=0.49\textwidth]{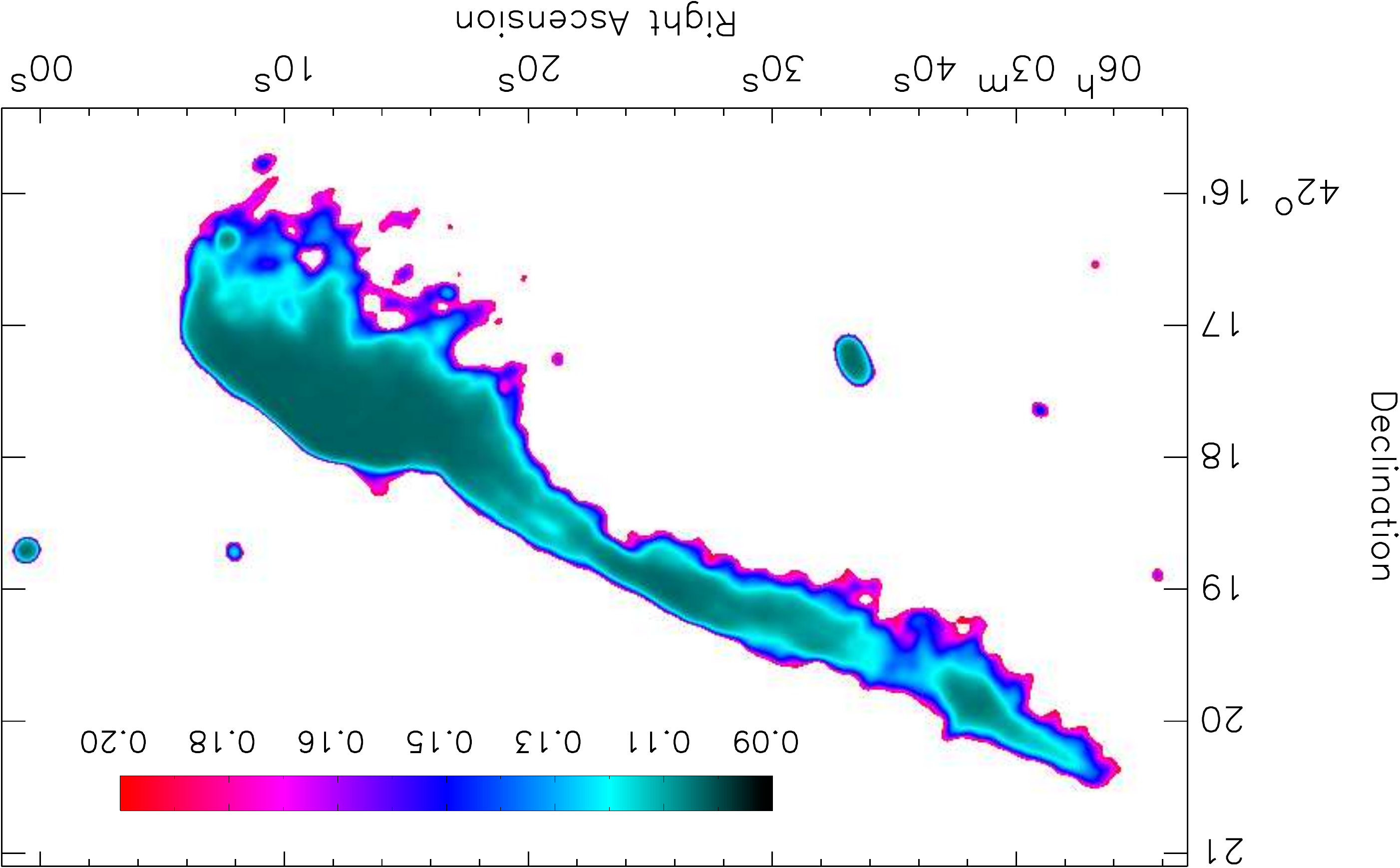}
\end{center}
\caption{Left: Spectral index map between 151 and 610~MHz at 6.5\arcsec~resolution. Black contours are drawn at levels of $[1,2,4,\ldots] \times 1.25$~mJy~beam$^{-1}$ and are from the 151~MHz  image. Pixels with values below $4.5\sigma_{\rm{rms}}$ were blanked. The circled source at the western end of the relic is discussed in Sect.~\ref{sec:re-acceleration}. Right: Corresponding spectral index error map. The errors are based on the individual r.m.s. noise levels in the two maps and using an absolute flux calibration uncertainty of 10\% at 150 and 610~MHz.}
\label{fig:spixhbagmrt}
\end{figure*}

\begin{figure*}[t]
\begin{center}
\includegraphics[angle =0, trim =0cm 14cm 0cm 0cm,width=0.325\textwidth]{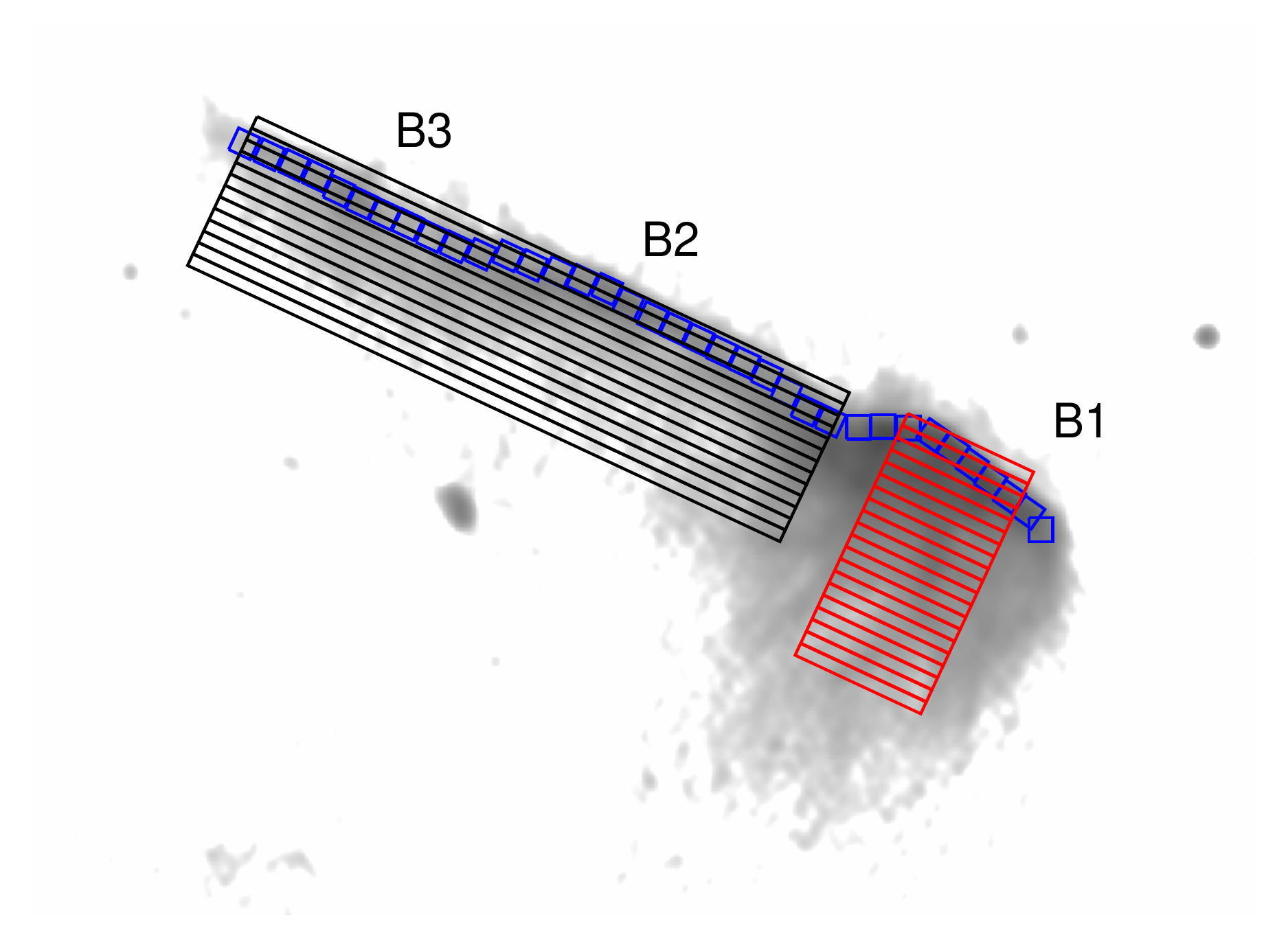}
\includegraphics[angle =180, trim =0cm 1.5cm 0cm 0cm,width=0.3\textwidth]{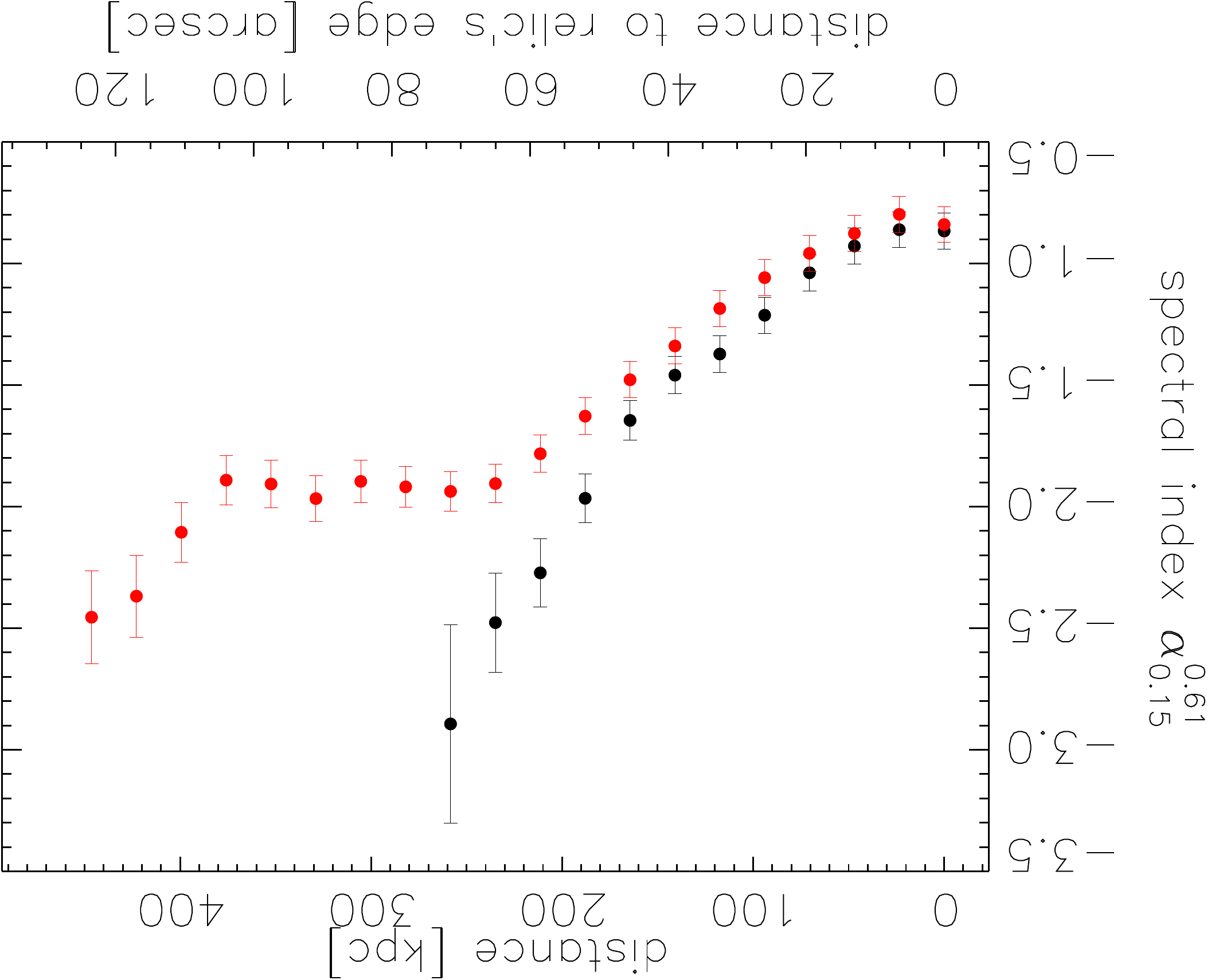}
\includegraphics[angle =180, trim =0cm 0cm 0cm 0cm,width=0.343\textwidth]{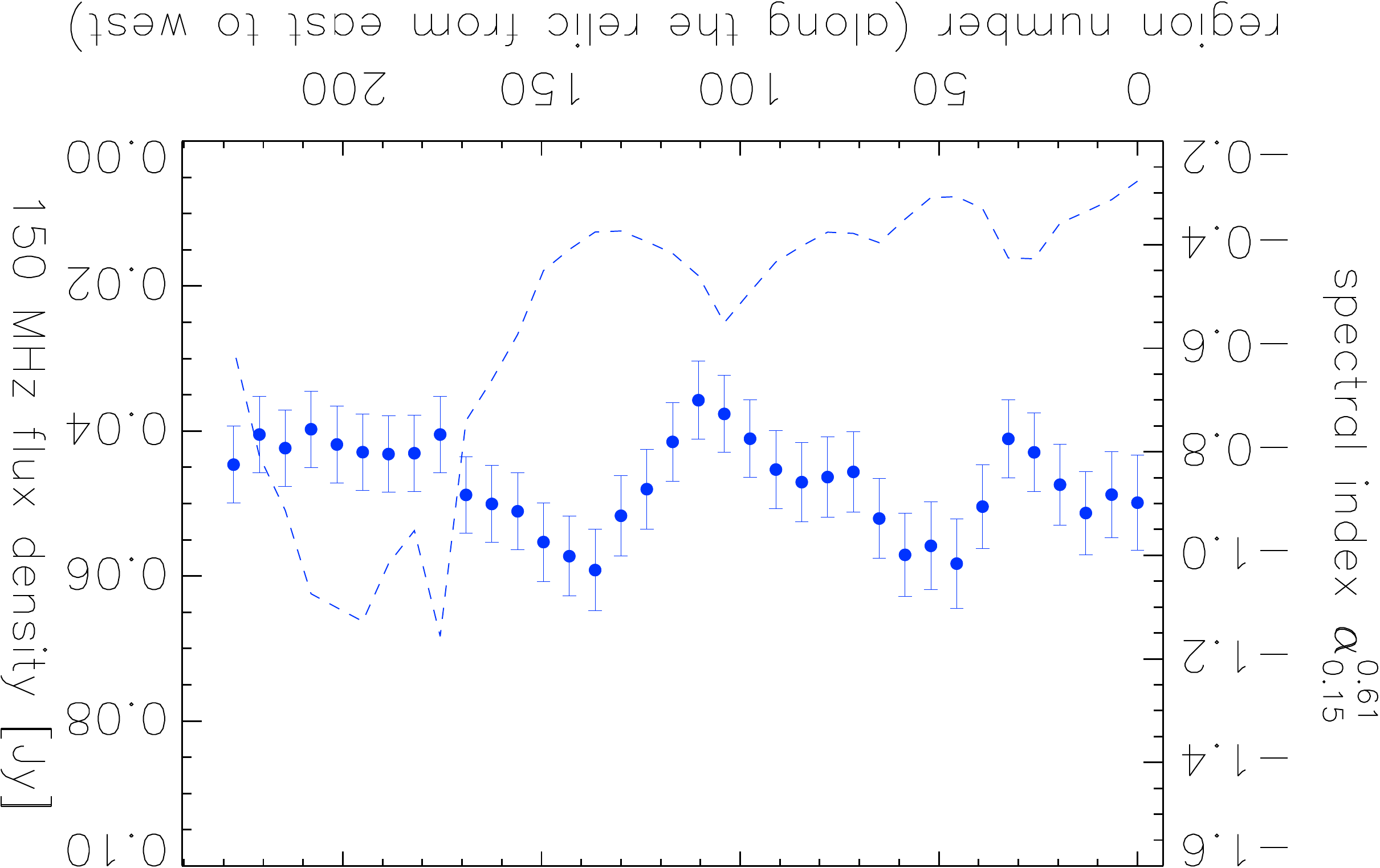}
\end{center}
\caption{Left: Regions where the spectral indices were extracted (displayed in the middle and right panel) on top of a HBA 150~MHz image (used to make Figure~\ref{fig:spixhbagmrt}). Middle: The spectral index between 150 and 610~MHz across relic B1, from north to south, is shown with red points. The black points cover the spectral index for the B2 and B3 part of the relic. The regions have a width of 6.5\arcsec. Right: The blue points trace the spectral index along the northern edge of relic B, with the distance increasing from east to west. The blue regions have sizes of 12\arcsec. The dashed blue line traces the LOFAR 150~MHz flux density along the relic, corresponding to the spectral indices shown in the same color.}
\label{fig:spixHRwedge}
\end{figure*}

The high resolution LOFAR observations allow us, for the first time, to map the low-frequency 150--610~MHz spectral index at $<10\arcsec$~resolution. To make the spectral index maps, we employed uniform weighting, used common inner uv-range cuts, and convolved the final images to the same resolution. The 6.5\arcsec~resolution 150--610~MHz spectral index map is shown in Figure~\ref{fig:spixhbagmrt}. We blanked pixels with values less than $4.5\sigma_{\rm{rms}}$ in both the 150 and 610~MHz maps. We also computed spectral profiles across the relic, see Figure~\ref{fig:spixHRwedge}. 

The spectral index map and profiles of relic B display a clear north-south spectral index gradient, with steepening from $\alpha=-0.8$ to $\alpha \lesssim -2.5$.  These numbers are consistent with our previous spectral index estimates ($\alpha \approx -0.6$ to $-0.7$) considering the difference of 10\% in the flux-scale between the GMRT and LOFAR 150~MHz maps. With the new results presented here, a spectral index as flat as $-0.6$ is however disfavored. In Figure~\ref{fig:spixHRwedge} we also display the variation of the spectral index along the northern edge of the relic from east to west (a distance of 1.9~Mpc).  The spectral index varies between $-0.7\pm0.1$ and $-1.0\pm0.1$, with the brighter parts corresponding to flatter spectra.

{The spectral index also steepens along the ``streams" of emission that originate from B1. However, at their southern ends we find that the spectral indices are somewhat flatter than the surrounding emission, by about $0.3$ to $0.5$ units (see Figure~\ref{fig:spixhbagmrt}).  The streams might be the result of stretched magnetic field lines, allowing electrons to diffuse more efficiently along these field lines. Also stronger magnetic fields, combined with a diffusion time smaller than the electron cooling time, would ``illuminate'' lower energy electrons giving a flatter spectral index.}

\begin{figure*}[t]
\begin{center}
\includegraphics[angle =180, trim =0cm 0cm 0cm 0cm,width=0.49\textwidth]{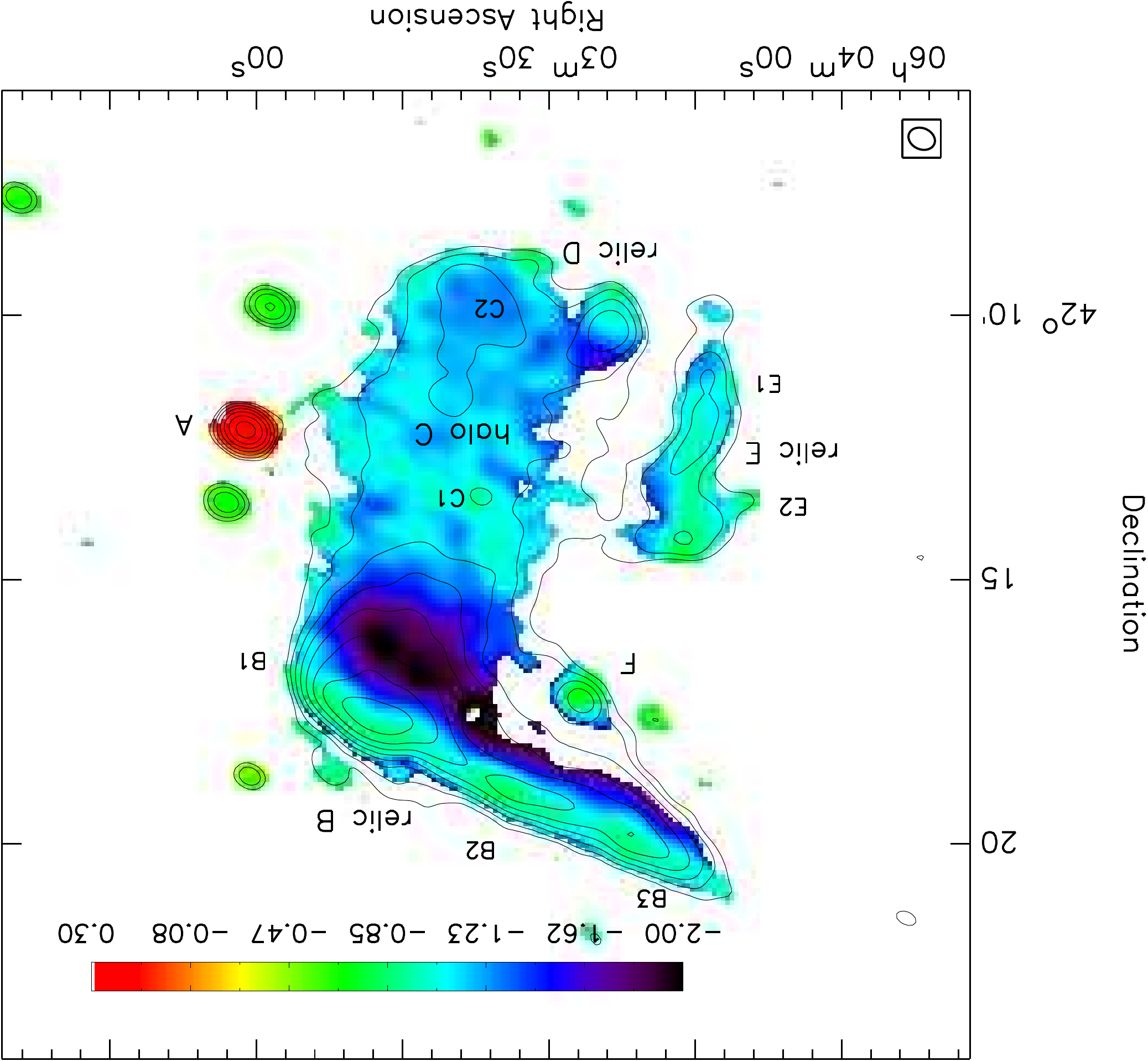}
\includegraphics[angle =180, trim =0cm 0cm 0cm 0cm,width=0.49\textwidth]{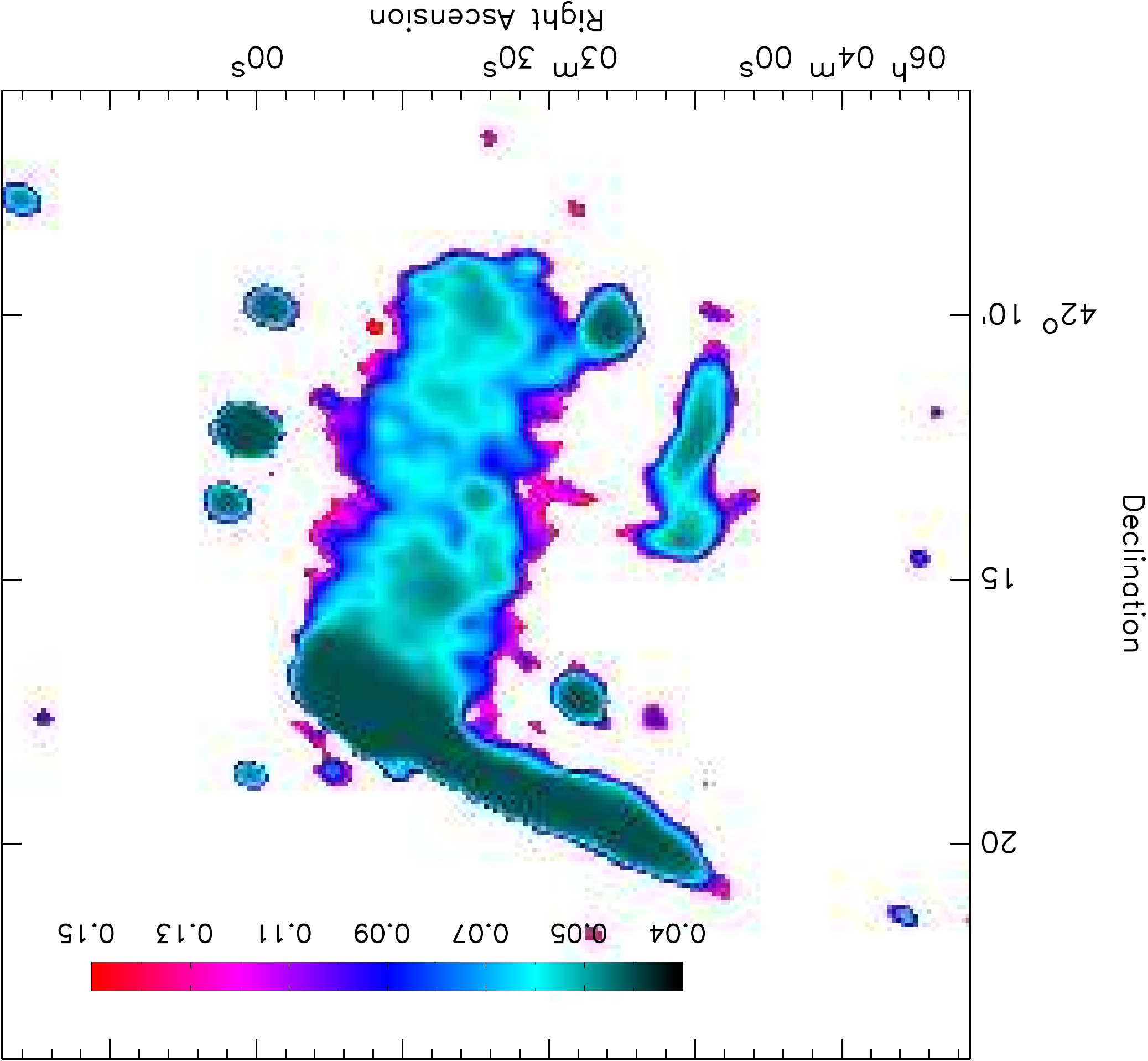}
\end{center}
\caption{Left: Spectral index map between 0.151 and 1.5~GHz at $31\arcsec\times24\arcsec$~resolution. Sources are labelled as in Figure~\ref{fig:hbaimage}). Black contours are drawn at levels of $[1,2,4,\ldots] \times 2$~mJy~beam$^{-1}$ and are from the 0.151~GHz image. Pixels with values below $3.5\sigma_{\rm{rms}}$ were blanked. Right: Corresponding spectral index error map. The errors are based on the individual r.m.s. noise levels in the two maps and using an absolute flux calibration uncertainty of 10\% at 150~MHz and 2.5\% at 1.5~GHz.}
\label{fig:spixhbajvla}
\end{figure*}

\begin{figure*}[t]
\begin{center}
\includegraphics[angle =0, trim =0cm 14cm 0cm 0cm,width=0.49\textwidth]{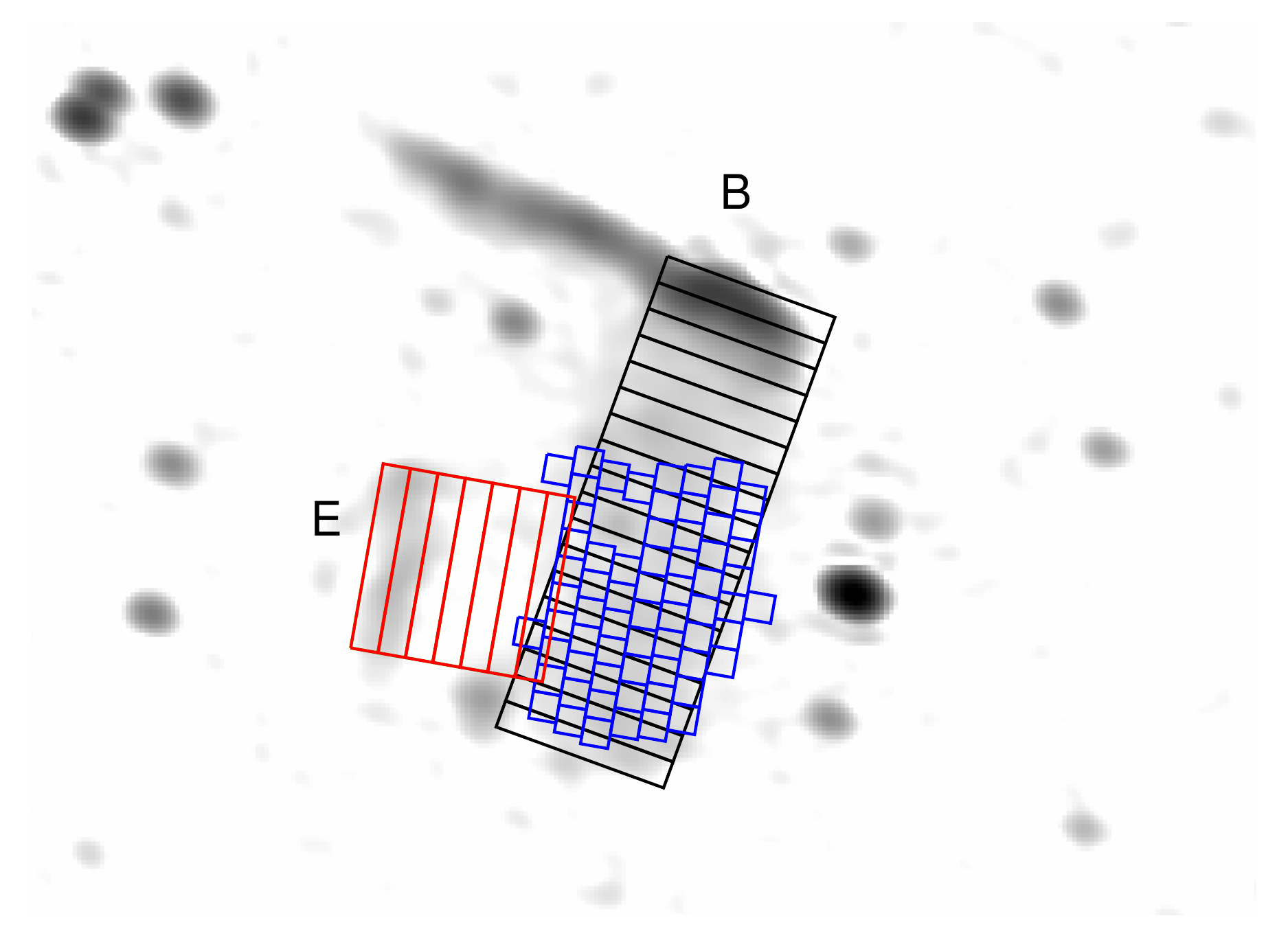}
\includegraphics[angle =180, trim =0cm 0cm 0cm 0cm,width=0.49\textwidth]{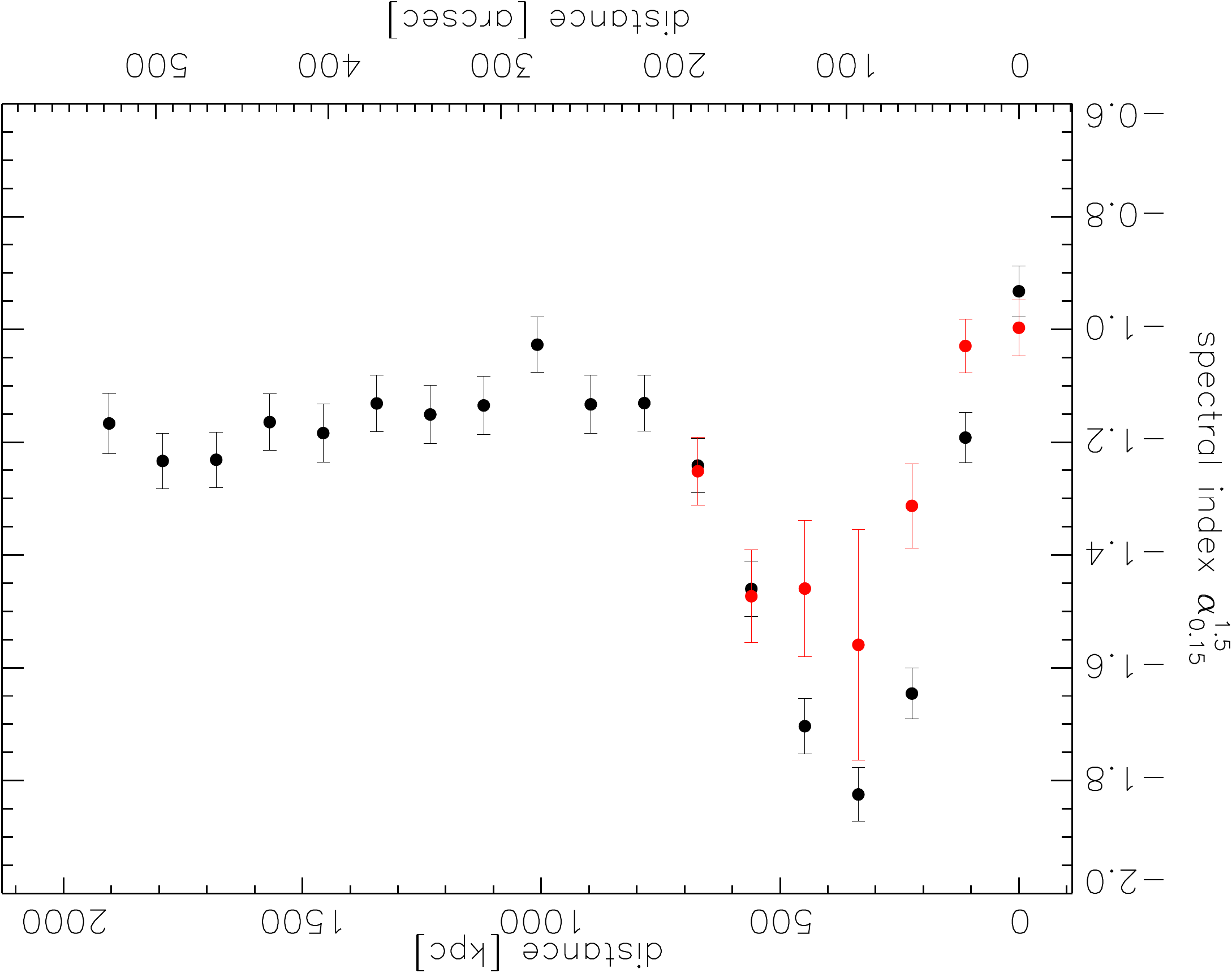}
\end{center}
\caption{Left: Regions where spectral indices were extracted are shown on top of the VLA 1--2~GHz image. The spectral indices for the red and black regions are shown in the left right panel. The spectra for the blue regions are shown in Figure~\ref{fig:spixhalo}. We excluded the regions affected by a compact source located at RA 06$^{\rm{h}}$03$^{\rm{m}}$23$^{\rm{s}}$, Dec 42\degr13\arcmin26\arcsec. Right: The spectral index between 0.15 and 1.5~GHz across the E radio relic is shown in red (from east to west). The spectral index across relic B1 and the radio halo, from north to south, is shown in black. The regions have a width of 31\arcsec.}
\label{fig:spixLRwedge}
\end{figure*}

We also created a ``low-resolution'' ($31\arcsec\times24\arcsec$) spectral index map by combining the LOFAR with the VLA 1--2~GHz D-array data (Figure~\ref{fig:spixhbajvla}). Pixels with values below $3.5\sigma_{\rm{rms}}$ were blanked. In this  ``low-resolution'' spectral index map,  we detect a spectral index gradient across source D, with $\alpha$ steepening from $-0.9$ to $-1.4$ from south to north. We again extracted spectral profiles in different regions (Figure~\ref{fig:spixLRwedge}). 
 We find a similar north-south trend across relic B1 as before, although in this case $\alpha$ only steepens to about $-1.8$ because of the limited spatial resolution and neighboring regions with flatter spectrum to the north and south. Similarly, the spectral index at the northern edge of relic B1 steepens to $-0.9$ due to mixing of emission from the steeper spectrum regions just to the south.  For the brightest part of relic E, the spectral index is $-1.0\pm0.1$. The spectrum then steepens towards the west to $\alpha =-1.6 \pm 0.2$, at the interface between relic E and radio halo C.  This is a similar trend as seen for relic B1.

After steepening across relic B1 from north to south, the spectrum flattens and remains relatively uniform across radio halo C. Going from C1 towards C2, the spectral index only steepens slightly from $-1.1$ to $-1.2$ (Figure~\ref{fig:spixLRwedge}). Across almost  the entire radio halo, the spectral index uncertainties are smaller than 0.1. This makes it the best radio halo spectral map constructed so far \citep[compare for example to the maps in, e.g.,][]{2014A&A...561A..52V,2007A&A...467..943O,2004A&A...423..111F}. In this case, the large spectral baseline, together with a deep low-frequency map, reduces the spectral index errors by a factor $\gtrsim 2$. Additionally, it increases the number of independent regions where the spectral index can be measured accurately (i.e., a spectral index uncertainty $< 0.1$) by about an order of magnitude. In Sect.~\ref{sec:spixhalo}, we discuss the radio halo spectral index in more detail.

\subsection{Chandra image}

\begin{figure*}[t!]
\begin{center}
\includegraphics[angle =180, trim =0cm 0cm 0cm 0cm,width=0.75\textwidth]{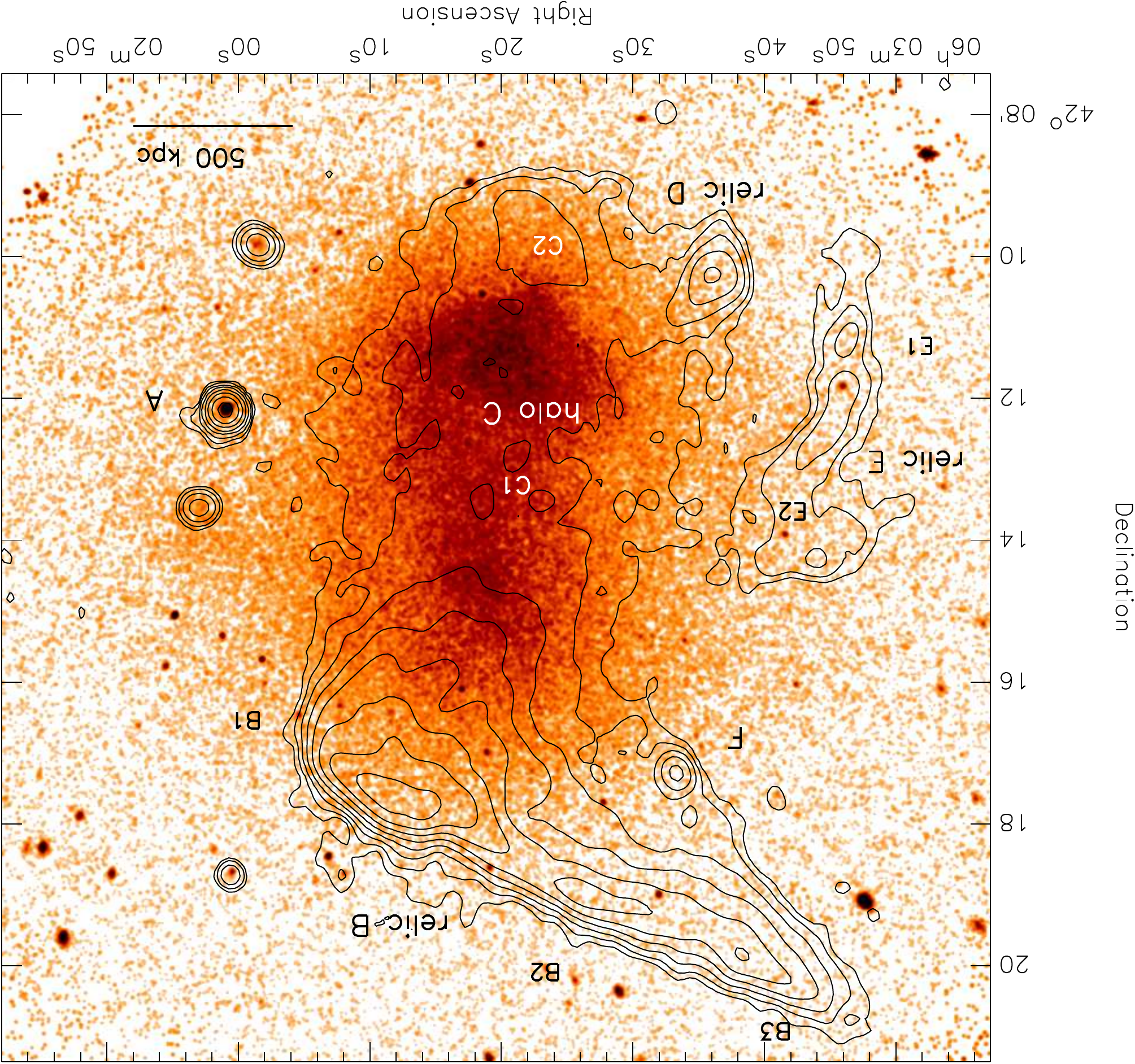}
\end{center}
\caption{{\it Chandra} ACIS-I 0.5--2.0~keV image (bin=2) of the Toothbrush cluster. The image is smoothed with a Gaussian with a FWHM of 3\arcsec. Radio contours from the LOFAR image with a resolution of 17.7\arcsec~(right panel Figure~\ref{fig:hbaimagelowres}) are overlaid. Radio sources are labelled as in Figure~\ref{fig:hbaimage}. The X-ray image shows two
main concentrations, aligned N-S. The southern structure is better defined and has a bullet-like shape (see also Figure~\ref{fig:secimage}).}
\label{fig:chandra}
\end{figure*}

\begin{figure}[t!]
\begin{center}
\includegraphics[angle =0, trim =0cm 0cm 0cm 0cm,width=0.49\textwidth]{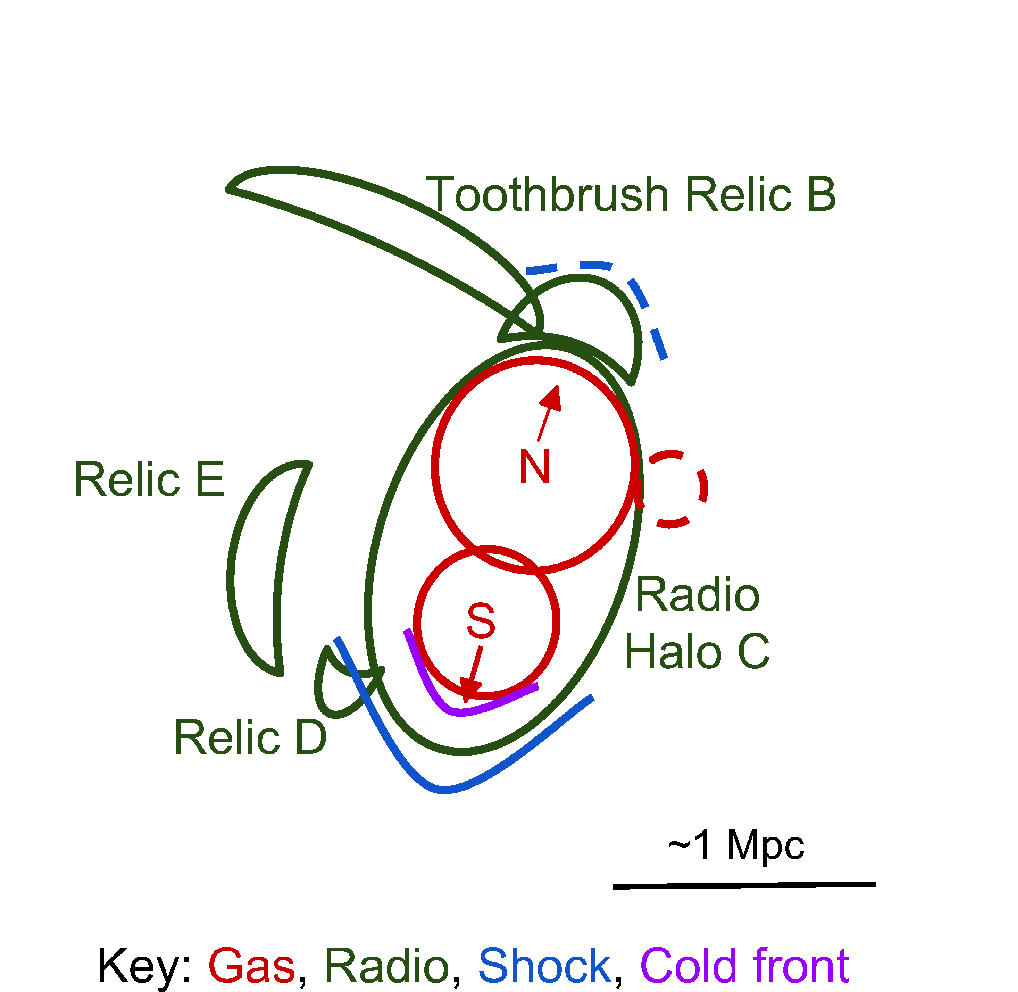}
\end{center}
\caption{Schematic overview of the radio and X-ray components detected in the cluster. The dashed lines shows features whose identification is less certain.}
\label{fig:cartoon}
\end{figure}

The {\it Chandra} 0.5---2.0~keV image of the cluster is displayed in Figure~\ref{fig:chandra}. It shows extended north-south X-ray emission consisting of two main concentrations which were also clearly seen in previously published {\it XMM-Newton} images  \citep{2013MNRAS.433..812O}. The southern subcluster is  better defined and brighter than the northern one.  The southern subcluster displays a bullet-like shape with a surface brightness edge on its southern side. Further to the south of this ``bullet'',  the {\it Chandra} image reveals a bow-shaped surface brightness edge. The edge coincides with the southern edge of the radio halo. We discuss the nature of these edges in Sect.~\ref{sec:edges}. To the west of the two main subclusters, about 1\arcmin~north of source~ A, we detect a smaller substructure (see also Figure~\ref{fig:secimage}). A schematic overview of the cluster is presented in Figure~\ref{fig:cartoon}.

\subsection{X-ray temperature map}
\label{sec:Xraytemperatures}

To create a temperature map of the cluster, we divided the cluster into individual regions whose edges follow the surface brightness contours using {\tt contbin}  \citep{2006MNRAS.371..829S}. This method is particularly useful for clusters where the surface brightness distribution has edges associated with density discontinuities. We required that each spectrum contains at least 5,000 counts per region in the 0.5--7.0~keV band. Compact sources were detected in the 0.5--7.0~keV  band with the {\tt CIAO} task {\tt wavdetect} using scales of 1, 2, 4, 8 pixels and {cutting at the $3\sigma$ level}. We manually inspected the output and added a couple more sources that were identified by eye. The resulting spectra were then fitted with {\tt XSPEC} \citep[v12.8.2,][]{1996ASPC..101...17A}. The regions covered by the compact sources were excluded from the spectra.  

\begin{figure*}[t!]
\begin{center}
\includegraphics[angle =180, trim =0cm 0cm 0cm 0cm,width=0.75\textwidth]{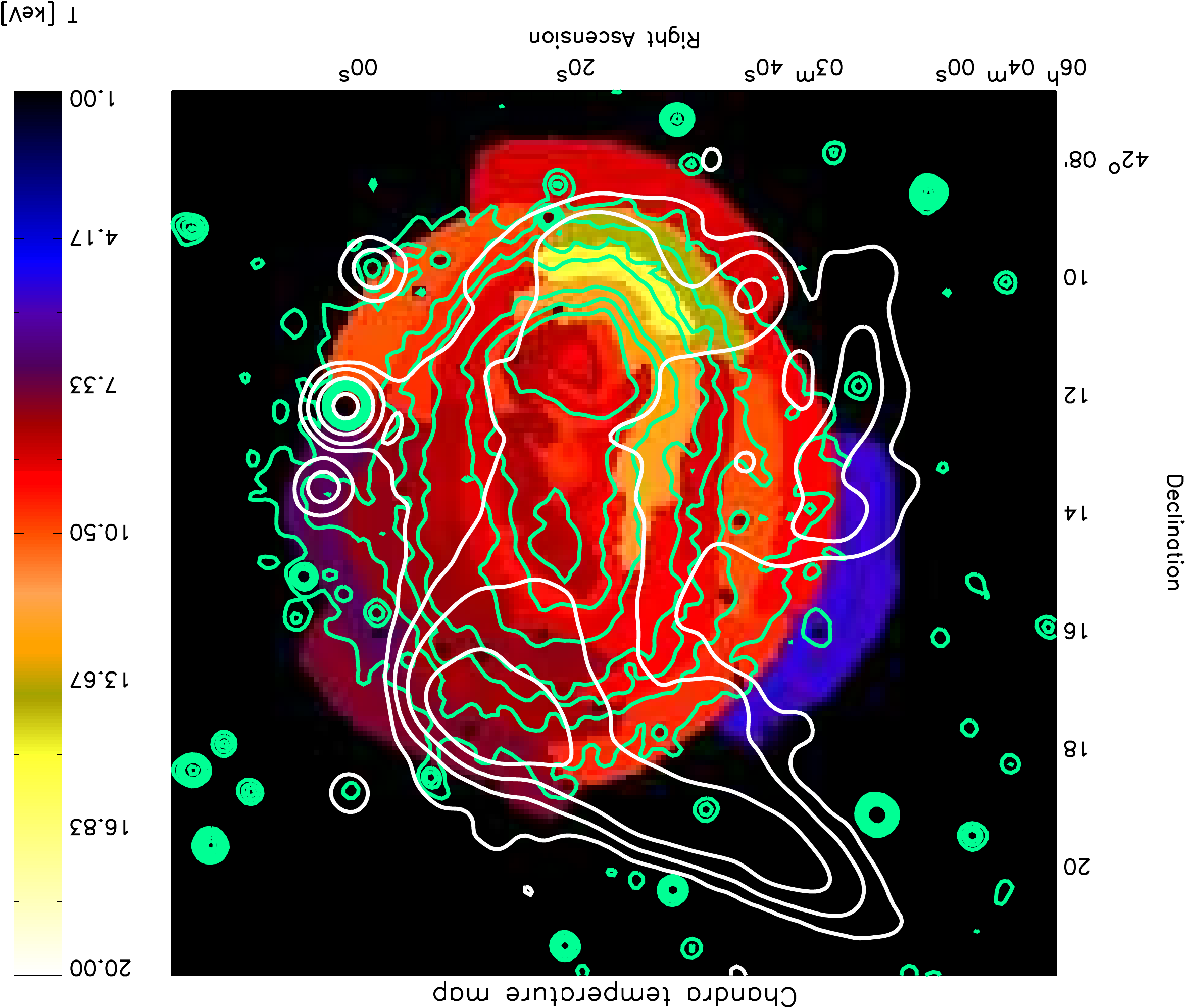}
\end{center}
\caption{X-ray temperature map of the cluster. Each region was required to contain at least 5,000 net counts in the 0.5--7.0~keV band. X-ray contours (green) and radio contours (white, Figure~\ref{fig:hbaimagelowres} left panel) are overlaid. Black colored regions are blanked, either due to compact sources or because of an insufficient number of ICM counts. }
\label{fig:tmap}
\end{figure*}

ICM spectra were fitted with an absorbed thermal emission model ({\tt phabs * APEC}). The metallicity  was fixed to a value of $0.3$~$Z_\odot$ using the abundance table of \cite{1989GeCoA..53..197A}. For the redshift we took the default value of $z=0.225$ \citep{2012A&A...546A.124V}.  The total Galactic \rm{H}~\rm{I} column density was fixed to a value of $N_{\rm{H}} = 0.214\times 10^{22}$~cm$^{-2}$ \citep[the weighted average $N_{\rm{H}}$ from the Leiden/Argentine/Bonn (LAB) survey,][]{2005A&A...440..775K}. The resulting temperature map is shown in Figure~\ref{fig:tmap}. The uncertainties on the temperatures are shown in Figure~\ref{fig:tmaperrs}.

For most of the cluster regions, we measure temperatures of 7--10~keV. The main deviation from that is a region of hot gas located to the south of the southern subcluster's core. The temperature in this region is  14--15~keV, with the $1\sigma$ {statistical uncertainties in the temperature of about} $-2$~keV and $+3$~keV. The presence of this hot gas was also hinted at by the {\it XMM-Newton} observations \citep{2013MNRAS.433..812O}. We extracted the spectra for the cores of the two main subclusters; these two regions are indicated in Figure~\ref{fig:secimage}. Using the same fitting parameters as for the temperature map, we find temperatures of $8.43^{+0.26}_{-0.25}$ and $9.00^{+0.28}_{-0.28}$~keV for the northern (\textit{coreN}) and southern (\textit{coreS}) subclusters, respectively.

\section{Characterization of surface brightness edges} 
\label{sec:edges}
A visual inspection of the {\it Chandra} image reveals two edges in the surface  brightness distribution in the southern subcluster. One of the surface brightness edges is located at the southern boundary of the radio halo, while the other one seems to be a contact discontinuity between the southern core and the hotter ICM in the cluster outskirts, see Figure~\ref{fig:chandra}. The locations of these discontinuities are indicated in Figures~\ref{fig:cartoon} and~\ref{fig:secimage}.

We used {\tt PROFFIT} \citep{2011A&A...526A..79E} to extract and fit the surface brightness profiles. The areas covered by the compact sources were excluded from the  fitting (see Sect.~\ref{sec:Xraytemperatures}). We modeled the surface brightness profiles across the edges assuming underlying broken power-law density models, plus a constant ($S_{\rm{back}}$) to optionally include a background level.  The emissivity is assumed to be proportional to the density squared and this model is then integrated along the line of sight assuming spherical symmetry. The sectors for which the fitting was performed are indicated in Figure~\ref{fig:secimage} (see also Table~\ref{tab:sectorfits}). 
For the inner ``edge'' of the southern subcluster's core we considered two different sectors (\textit{coreS\_east, coreS\_west}) as the edge has a peculiar right-angled shape with a small apex. For the binning of photons for the partial annuli, we required a SNR of 5 per bin. 

The broken power-law model, which can be used to represent a density jump associated with a shock or cold front, is given by
\begin{equation}
  n(r)=%
  \begin{cases}
    \left(n_{2}/n_{1}\right) n_0 \left(\frac{r}{r_{\rm{edge}}} \right)^{a_2}  \mbox{ ,} &\text{$r \le r_{\rm{edge}}$} \\
    \\
    n_0 \left(\frac{r}{r_{\rm{edge}}} \right)^{a_1}  \mbox{ ,}&\text{$r > r_{\rm{edge}}$}   \mbox{ .}
  \end{cases}
  \label{eq:proffit}
\end{equation}
The subscripts 1 and 2 denote the upstream and downstream regions, respectively. The parameter $n_0$ is a normalization factor, $a_1$ and $a_2$ define the slopes of the power-laws, and $r_{\rm{edge}}$ the position of the jump. The parameter $n_{2}/n_{1} \equiv C $ represents the electron density jump. In the case of a shock,  $n_{2}/n_{1}$ is related to the Mach ($\mathcal{M}$) number via the Rankine-Hugoniot  relation 
\begin{equation}
{\cal M}=\left[\frac{2 C}{\gamma + 1 - C(\gamma -1)}\right]^{1/2}   \mbox{ ,}
\label{eq:machne}
\end{equation}
\noindent where $\gamma$ is the adiabatic index of the gas. In this work, we assume $\gamma=5/3$.

The modeling was done assuming a fixed background level of zero ($S_{\rm{back}} = 0$) because the {\it Chandra} instrumental and sky background components are already subtracted.

To determine the nature of the density jumps (i.e., shocks or cold fronts), we also need to measure temperatures on both sides of $r_{\rm{edge}}$. For that we defined regions on each side of $r_{\rm{edge}}$ and extracted the spectra from the {\it Chandra} data.  The locations of these regions are indicated in Figure~\ref{fig:secimage}. We fit the temperatures in the same way as described in Sect.~\ref{sec:Xraytemperatures}. What we fit is thus the projected temperature, which is an approximation of the real 3D temperature. {We note that the regions where the temperatures are extracted are quite large. We did not attempt to correct for possible temperature gradients or variations within these bins.}

For a cold front, the pressure is the same on both sides of $r_{\rm{edge}}$. For a shock and assuming $\gamma=5/3$, the Rankine-Hugoniot jump conditions relate the upstream ($T_1$) and downstream temperature ($T_2$) to the Mach number \citep[e.g.,][]{2010ApJ...715.1143F}
\begin{equation}
{\cal M}=   \frac{ \left( 8 \frac{T_2}{T_1} - 7 \right) + \left[ \left( 8 \frac{T_2}{T_1} - 7 \right)^2 + 15 \right]^{1/2} }{5}    \mbox{ .}
\label{eq:machT}
\end{equation}

\subsection{Southern sector}
\label{sec:southsector}
For the southern discontinuity (at the edge of the radio halo), we find that the broken power-law model provides a good fit to the data, with $C = 1.57^{+0.09}_{-0.08}$, see also Table~\ref{tab:sectorfits}\footnote{We repeated the fit, but not subtracting the sky background and leaving  $S_{\rm{back}}$ as a free parameter.  The value of  $C=1.65^{+0.12}_{-0.11}$ we obtained in this way is consistent with our earlier result.}. The profile and resulting best-fitting model are shown in {the right panel of} Figure~\ref{fig:profileshocks}.

For the temperatures on both sides of the surface brightness edge, we measure $T_{1S} = 7.8^{+1.6}_{-1.4}$~keV and $T_{2S} = 14.5^{+2.9}_{-1.8}$~keV. Using Eq.~\ref{eq:machT},  we find $\mathcal{M} = 1.8^{+0.5}_{-0.3}$. Taking the density jump and Eq.~\ref{eq:machne} we obtain $\mathcal{M} = 1.39^{+0.06}_{-0.06}$.  Thus the results from the temperature and density jump indicate that the southern brightness edge traces a shock. 

{We also divided the southern sector into three equal parts, with opening angles of 40\degr~to search for variations of the Mach number along the shock front. From east to west we obtain $\mathcal{M} = 1.16^{+0.11}_{-0.07}$ (210\degr--250\degr),  $\mathcal{M} = 1.57^{+0.16}_{-0.13}$ (250\degr--290\degr), and $\mathcal{M} = 1.43^{+0.09}_{-0.08}$ (290\degr--330\degr)}, based on the density jump determined from the surface brightness profile. This result indicates that the  eastern part of the southern shock is weaker.


\begin{table*}
\begin{center}
\caption{Surface brightness profiles}
\begin{tabular}{llllll}
\hline
\hline
& south & coreS\_east & coreS\_west & north\\
\hline
sector center (RA, J2000)  		& 90.835417\degr 					&90.785958\degr  					&90.862000\degr				&90.853138\degr\\
sector center (DEC, J2000)		& 42.188797\degr 					&42.223506\degr 					&42.239089\degr 				&42.225425\degr\\
opening angle    		&210\degr--330\degr  				&200\degr--230\degr  				&280\degr--310\degr  			&40\degr--75\degr\\ 
fitting range			&1.2\arcmin--7.0\arcmin\				&3.0\arcmin--5.3\arcmin				&3.2\arcmin--5.2\arcmin			&2.6\arcmin--10.0\arcmin\\	
${n_2}/{n_1}$			& $1.57^{+0.09}_{-0.08}$ 			&$1.35^{+0.05}_{-0.05}$ 			&$1.46^{+0.05}_{-0.05}$		&$1.26^{+0.14}_{-0.10}$\\
$r_{\rm{edge}}$		& $2.45\arcmin^{+0.02}_{ -0.02}$ 	&$4.09\arcmin^{+0.03}_{-0.02}$		&$4.12\arcmin^{+0.03}_{-0.02}$	&$5.31\arcmin^{+0.17}_{-0.19}$\\
$\chi^2/{\rm{d.o.f.}}$		& {68.4}/{76}						& {62.0}/{48}						& {56.5}/{41}					&{68.7}/{76}\\
\hline
\hline
\end{tabular}
\label{tab:sectorfits}
\end{center}
\end{table*}

\subsection{Southern core}

For the southern subcluster's eastern sector (\textit{coreS\_east}), we obtained $C=1.37^{+0.05}_{-0.05}$. For the western sector (\textit{coreS\_west}), we obtained $C=1.46^{+0.05}_{-0.05}$, which indicates very similar density jumps (see Figure~\ref{fig:profilecold} for the profiles). The reduced $\chi^2$ of 1.3--1.4 (Table~\ref{tab:sectorfits}) means the fits are reasonable but not perfect, which is not surprising given the assumption of spherical symmetry and the power-law approximation to the inner density slope. 

For the temperatures on both sides of the edge, we find  $T_{1} = 11.4^{+1.2}_{-0.9}$~keV and $T_{2} = 9.0^{+0.4}_{-0.4}$~keV. The subscripts 1 and 2  refer  to the regions indicated in Figure~\ref{fig:secimage}. Here, we combined the eastern and western regions for the temperature measurements,  motivated by the very similar density jumps measured for the eastern and western sectors (\textit{coreS\_east} and \textit{coreS\_west}).  
Taking the temperature and density jumps, we find that the pressures on both sides of $r_{\rm{edge}}$ are consistent within their errors as indicated by the ratio $T_1/T_2$ and $n_2/n_1$. We therefore classify this surface brightness edge as a cold front. This combination of a merger related cold front and a shock ahead of it (Sect.~\ref{sec:southsector}), is very similar to what is seen in the Bullet Cluster \citep{2002ApJ...567L..27M}.

The distance between a shock and the blunt body producing the shock \citep[i.e., the remnant subcluster's core,][]{2001ApJ...551..160V} depends on the Mach number. Applying this to the Toothbrush cluster, the distance ratio between shock and cold front location ($\approx 2$) is consistent with our obtained Mach number \citep[eq.~11,][]{1994JGR....9917681F}.
This confirms the association between the southern subcluster's cold front and the shock ahead of it. We can also use the Mach angle ($\mu$) to compute the Mach number of the shock, via the relation $\sin{\left(\mu\right)} = \frac{1}{\cal M}$. From Figure~\ref{fig:chandra}, we estimate a Mach angle $\mu \approx 40 \degr$ which corresponds to $\mathcal{M} \approx 1.5$, {again} consistent with the previous results.

\subsection{Northern sector}

In addition to the three sectors in the southern part of the cluster, we also extracted a profile across the northern radio relic (B1) to search for a possible edge associated with the front of the relic, see Figure~\ref{fig:secimage}. The sector shape is defined by the curvature and shape of the radio relic. The resulting profile is shown in {left panel of} Figure~\ref{fig:profileshocks}. Interestingly, the profile indicates a change in the slope of the surface brightness profile around the location of the radio relic. We fitted a broken power-law density model to this profile and obtained $C=1.26^{+0.14}_{-0.10}$, with the model providing a good fit to the data. The position of the jump ($r_{\rm{edge}}$) is also in good agreement with the location of the front of the radio relic. However, at the 90\% confidence limit we cannot exclude the possibility of having a kink in the profile (i.e., $C=1$). Better statistics will be needed to determine if a jump is really required.

For the northern partial annuli across relic B1, we measure $T_{1N} = 8.3^{+3.2}_{-2.4}$~keV and $T_{2N} = 8.2^{+0.7}_{-0.9}$~keV. Clearly, the error on the putative pre-shock temperature is too large to confirm or rule out a shock, which is expected given the low Mach number and the low count rate in the pre-shock region. Taking the $1\sigma$ lower and upper limit for the pre and post-shock temperatures, respectively, we obtain an upper limit on the Mach number of $1.5$.

Interestingly,  both {\it Chandra} and {\it XMM-Newton} data show that any shock that traces the  Toothbrush relic (the B1-part) must have a very low Mach number, otherwise the jump should have been detected unambiguously. To determine an upper limit on the Mach number, we refitted the {\it Chandra} profile keeping the location of the shock fixed to the value reported in Table~\ref{tab:sectorfits} (which coincides with the front of relic B1). From this we determined an upper limit of $C=1.7$ at the 90\% confidence level. 
Thus we find that the putative shock must have a Mach number of $\mathcal{M} \le 1.5$. This number is consistent with the upper limit we found from the temperature change across the relic.

\subsection{Comparison with XMM-Newton results}
We compare the above results with previously published  {\it XMM-Newton} data by \cite{2013MNRAS.433..812O}. {\it XMM-Newton} also found a shock in the southern part of the cluster. The Mach number measured by {\it Chandra} via the density jump is consistent with the $\mathcal{M} = 1.5 \pm 0.1$ reported by \citeauthor{2013MNRAS.433..812O}, but the actual position of the shock does not agree with the position found by {\it Chandra}. 
After a closer investigation, we conclude that small-scale substructure in the downstream region of the ICM appeared as an edge in XMM's surface brightness profile. Together with XMM's lower spatial resolution this caused the fitting procedure to pick up the wrong shock location. We also note that the sector in which the surface brightness profile is fitted is different from the one used by \cite{2013MNRAS.433..812O}. Since the edge is not clearly visible in the XMM image by eye there was no clear guidance on what sector to pick in this case. However, if we take the position of the shock found by {\it Chandra} as an initial guess and re-fit the XMM data, we obtain consistent results for $r_{\rm{edge}}$.

Turning to the northern subcluster, a less significant edge was reported to be located about 1\arcmin~to the north of relic B1 in the XMM image \citep[see figs.~8--11 from][]{2013MNRAS.433..812O}. At that location there is no jump visible in the {\it Chandra} profile. Given the rather low-significance ($<2\sigma$) of this jump its status remains unclear. We also note that the actual significance of this jump is lower than the fitting results suggest due to the look-elsewhere effect. 

\begin{figure*}[t!]
\begin{center}
\includegraphics[angle =0, trim =0cm 0cm 0cm 0cm,width=0.95\textwidth]{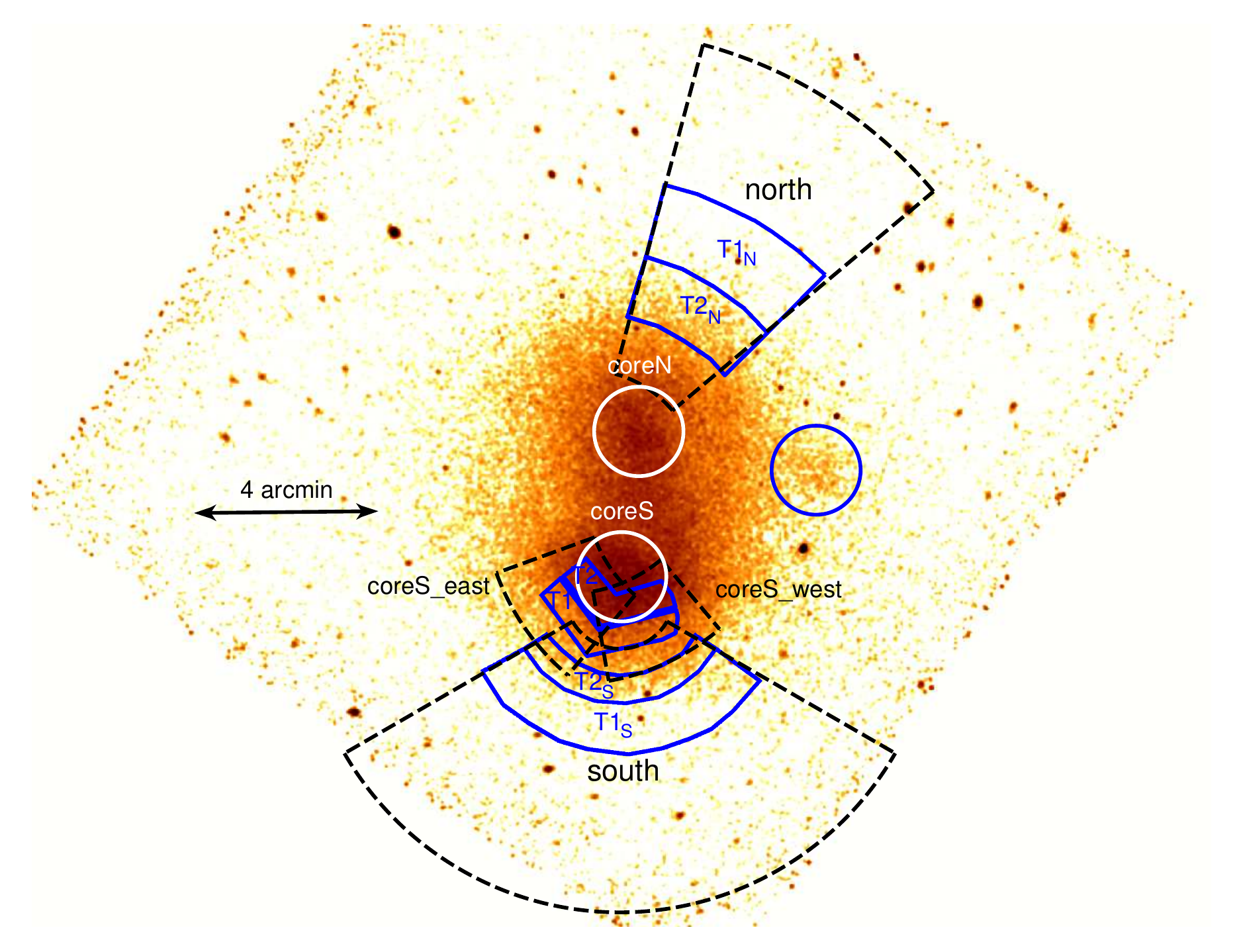}
\end{center}
\caption{Smoothed (4 pixels) {\it Chandra} 0.5--2.0~keV image showing the sectors (dashed) in which we extracted and fitted the surface brightness profile, see Table~\ref{tab:sectorfits}. The regions where the temperatures ($T_1$ and $T_2$) are extracted are shown in blue. The surface brightness edges are located at the boundary between the $T_1$ and $T_2$ regions. The blue circle represents the area where we extracted the X-ray luminosity of the western substructure. The white circles indicate the areas for which we extracted the temperatures for the cores of the northern ({\it coreN}) and southern ({\it coreS}) subclusters.}
\label{fig:secimage}
\end{figure*}

\begin{figure*}[t!]
\begin{center}
\includegraphics[angle =0, trim =0cm 0cm 0cm 0cm,width=0.49\textwidth]{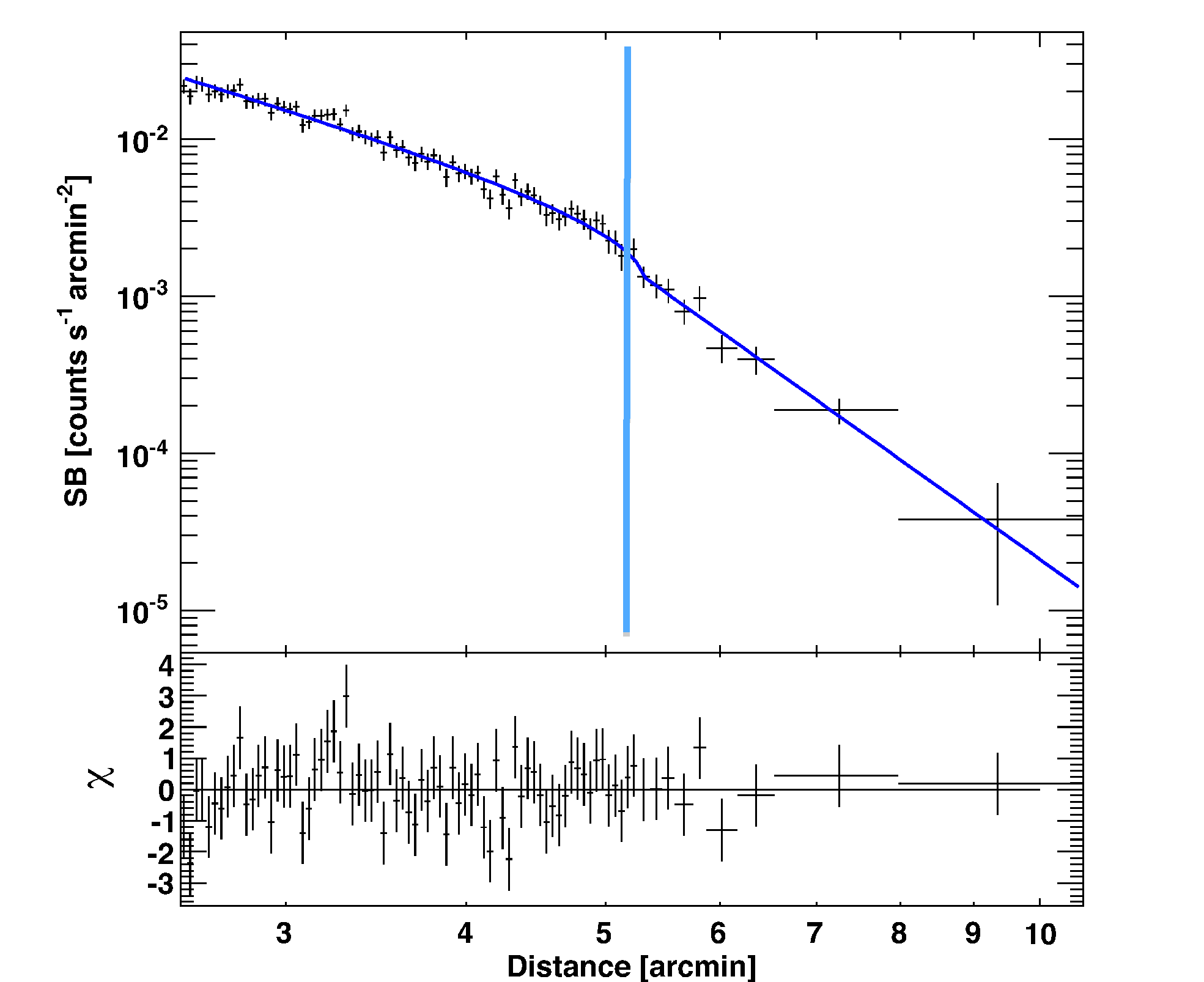}
\includegraphics[angle =0, trim =0cm 0cm 0cm 0cm,width=0.49\textwidth]{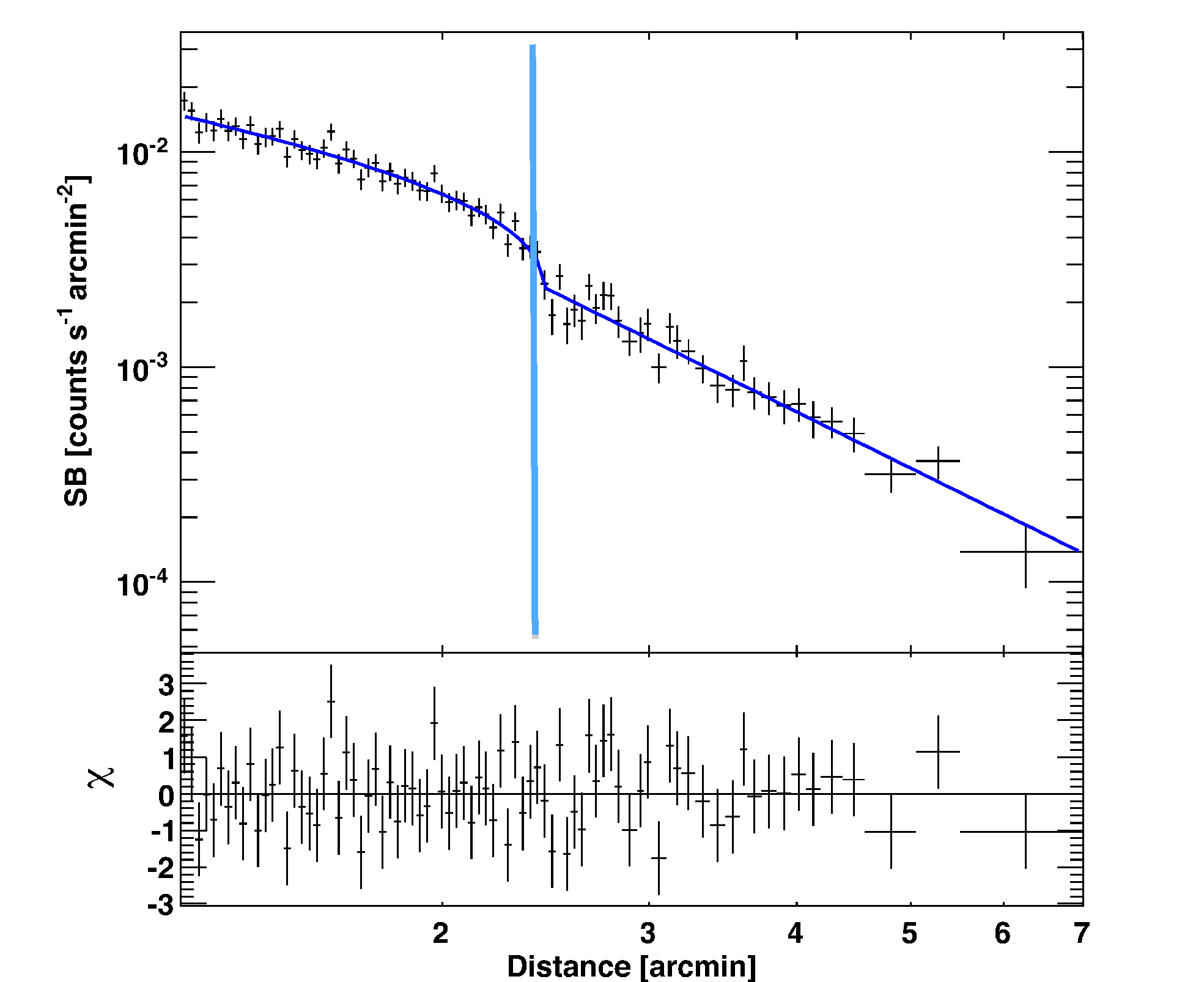}
\end{center}
\caption{Left: X-ray surface brightness profile across the bright part (B1) of the northern Toothbrush relic: Right: X-ray surface brightness profile in the southern sector. The blue lines show the best-fitting broken power-law density model (Eq.~\ref{eq:proffit}) and the vertical line the outer boundary of the radio relic or halo emission.\vspace{5mm}}
\label{fig:profileshocks}
\end{figure*}

\begin{figure*}[t!]
\begin{center}
\includegraphics[angle =0, trim =0cm 0cm 0cm 0cm,width=0.49\textwidth]{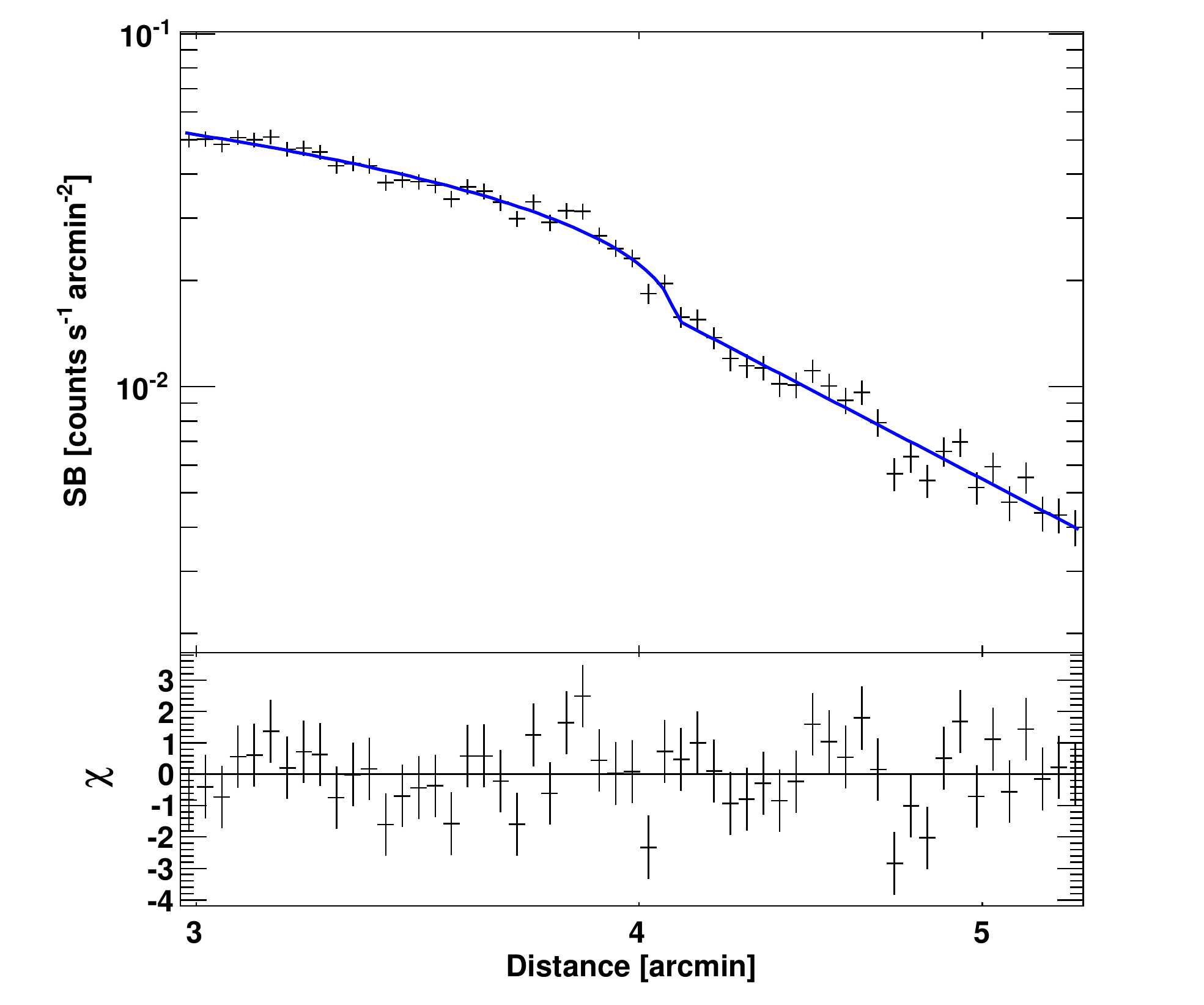}
\includegraphics[angle =0, trim =0cm 0cm 0cm 0cm,width=0.49\textwidth]{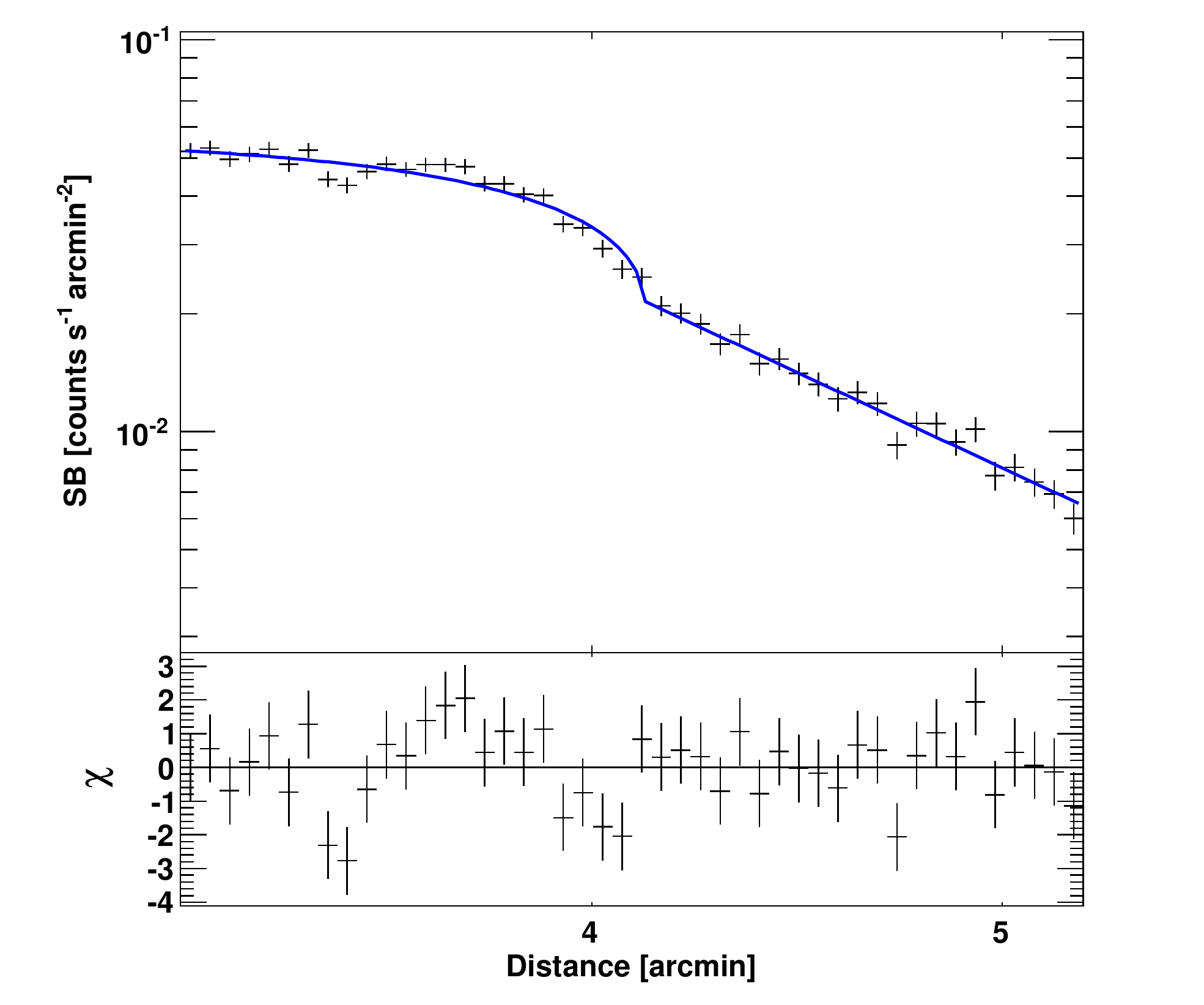}
\end{center}
\caption{Left: X-ray surface brightness profiles for the right-angled edges around the southern cluster's core. The profile for the eastern sector is shown in the left panel and the western one in the right panel, see Figure~\ref{fig:secimage}. The blue lines show the best-fitting broken power-law density model (Eq.~\ref{eq:proffit}). }
\label{fig:profilecold}
\end{figure*}

\cite{2013MNRAS.433..812O} did not find a shock at the front of relic B1, although a slightly different sector was used to extract the surface brightness profile. We refitted the XMM profile using the sector shown in Figure~\ref{fig:secimage} with a broken power-law model. The initial guesses for the fitting were taken from the best fit {\it Chandra} model (Figure~\ref{fig:profileshocks}). In this case, the fitted XMM profile was consistent with the results from {\it Chandra} and also revealed a change in the slope of the surface brightness profile at the front of relic B1.

We  conclude that both the {\it XMM-Newton} and {\it Chandra} data do not provide substantial evidence for a ``relic shock offset problem''. Instead, a change  in the slope of the surface brightness profile is observed at the expected location of a shock that would trace relic B1. The best-fitting broken power-law model indicates a small jump in the profile, corresponding to $C = 1.26^{+0.14}_{-0.10}$, at the location where the slope changes. However, the current {\it Chandra} data are not deep enough to fully exclude the possibility that this feature is just a change in the slope of the profile and not a jump.

\section{Discussion}

\label{sec:dicussions}
\subsection{Radio spectral index and shock Mach number for relic B1}
For a  
shock front, DSA (in the linear test particle regime) relates the shock Mach number to the injection spectral index $\alpha_{\rm{inj}}$ \citep[e.g.,][]{1987PhR...154....1B}
\begin{equation}
\alpha_{\rm{inj}} =  \frac{1}{2} - \frac{\mathcal{M}^2 +1} {\mathcal{M}^2 -1} \mbox{ .}
\label{eq:inj-mach}
\end{equation}
The injection spectral index is related to the energy spectrum of the electrons via $\alpha_{\rm{inj}} = -(\delta_{\rm{inj}}-1)/2$, with $dN/dE \propto E^{-\delta_{\rm{inj}}}$.
The volume integrated spectral index, $\alpha_{\rm{int}}$, is steeper by  0.5 units compared to $\alpha_{\rm{inj}}$ for a simple planar shock model where the electron cooling time is much shorter than the lifetime of the shock \citep{1969ocr..book.....G}. Recently, it has been shown that this approximation ($\alpha_{\rm{inj}} = \alpha_{\rm{int}} + 0.5$) does not hold for spherically expanding shocks with properties comparable to those found in clusters \citep{2015JKAS...48....9K,2015JKAS...48..155K}. Therefore, if possible, it is preferred to take the spectral index from spatially resolved maps to obtain $\alpha_{\rm{inj}}$ and avoid using $\alpha_{\rm{int}}+0.5$ in combination with Eq.~\ref{eq:inj-mach}. 

For the Toothbrush relic (source component B1, Figure~\ref{fig:hbaimage}) with $\alpha_{\rm{inj}}= -0.8 \pm 0.1$, we obtain $\mathcal{M} = 2.8^{+0.5}_{-0.3}$, which is not consistent with the limit of 1.5 set on the Mach number by the density jump. Working the other way around, i.e., starting with a Mach number of 1.5 (to take the most extreme value), we calculate $\alpha_{\rm{inj}} =-2.1$, 
which is in strong conflict with the radio spectral index maps. 

One could argue that the limit on the Mach number obtained from the density jump is not reliable, given the assumption of spherical symmetry and unknown radius of curvature. However, the Mach number derived via the temperature jump should be less sensitive to these effects. Given that we obtained a limit of about $\mathcal{M} \le 1.5$, from the temperature jump we conclude that it is unlikely that there could be a strong well-defined  $\mathcal{M} = 2.8$ shock hiding in the data.  

Recent PIC simulations indicate that particles are initially accelerated via SDA (also a Fermi I process) and not DSA \citep{2014ApJ...797...47G,2014ApJ...794..153G} at cluster shocks. However, these simulations suggest that the resulting  radio spectral index will be similar to that created by DSA. 

We therefore conclude that using the radio spectral index and Eq.~\ref{eq:inj-mach} leads to an overestimation of the X-ray-predicted Mach number for the Toothbrush relic, considering in particular that we took the most ``extreme scenario'', i.e., the steepest possible injection spectral index and the largest Mach number allowed by the X-ray surface brightness profile.

\subsection{Acceleration efficiency}
If the Mach number is indeed low, as suggested by the X-ray observations, another important consideration is the electron acceleration efficiency required to produce such a bright radio relic. The kinetic energy flux that becomes  available at the shock ($\Delta F_{\rm{KE}}$) 
is:
\begin{equation}
\Delta F_{\rm{KE}} = 0.5\rho_1 v_s^{3}  \left(1 - \frac{1}{C^2}\right)\mbox{ ,}
\end{equation}
where $v_{s}$ is the shock speed and $\rho_1$ is the upstream density. Part of this energy can be converted into electron acceleration and then
into synchrotron and IC emission.

\begin{figure}[t]
\begin{center}
\includegraphics[angle =0, trim =0cm 0cm 0cm 0cm,width=0.49\textwidth]{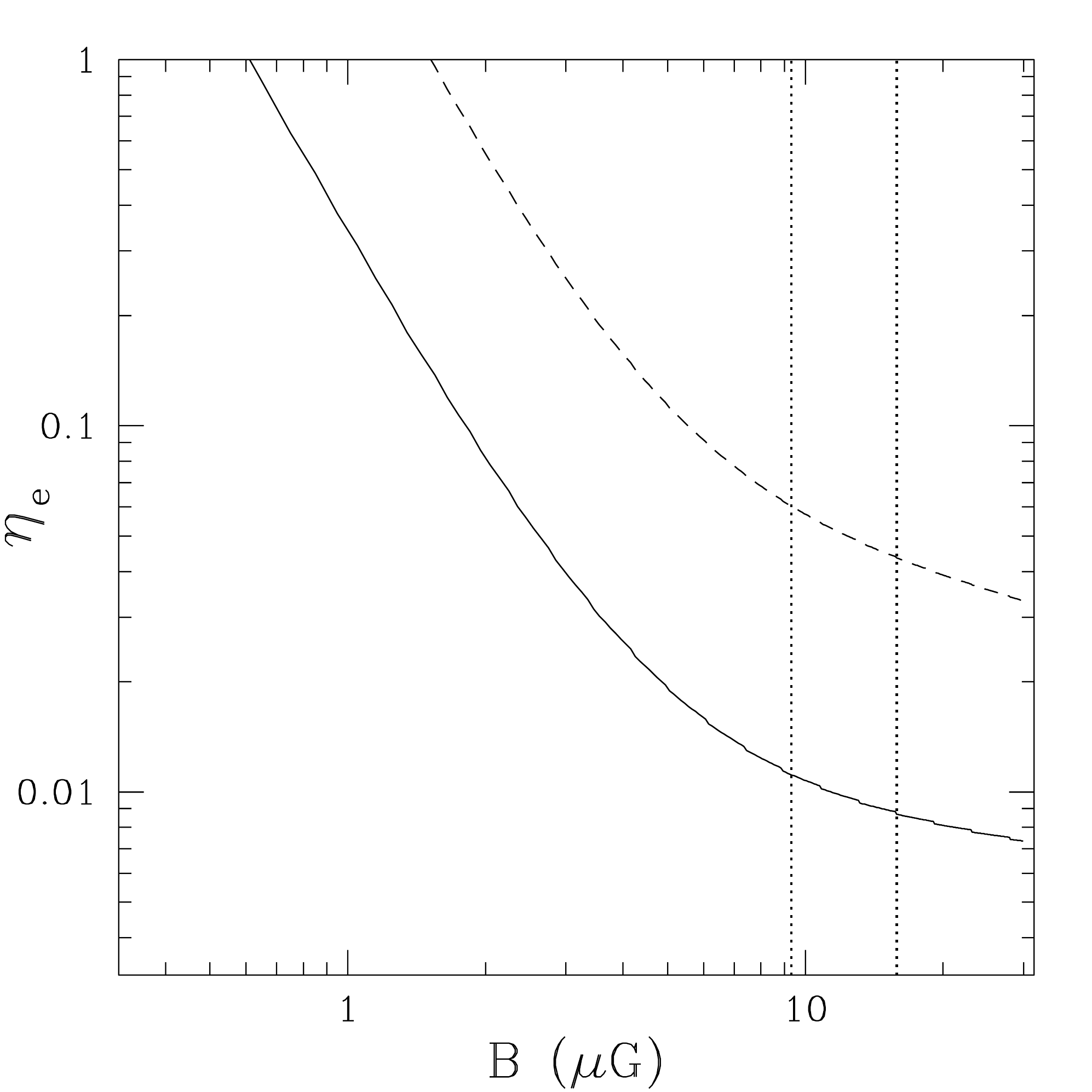}
\end{center}
\caption{The fraction of the kinetic energy flux at the shock ($\eta_e$) that needs to be channelled into the supra-thermal and relativistic electron components that is required to explain the observed flux density of relic B1. The solid line is for $\delta_{\rm{inj}}=2.2$ and the dashed line for   $\delta_{\rm{inj}}=2.6$. For any reasonable value of the magnetic field strength ($B \lesssim10$~$\mu$G) we find the  extreme requirement that more than several percent of the kinetic energy flux crossing the shock should be converted into the acceleration of relativistic electrons. The two vertical dotted lines indicate the magnetic field value for plasma $\beta$ values of 3 and 1.
}
\label{fig:etarelic}
\end{figure}

In Figure~\ref{fig:etarelic}, we show the efficiency of electron acceleration that
is necessary at the shock to explain the observed synchrotron luminosity \citep{2014IJMPD..2330007B} of the ``bright''
part of the relic, component B1. The efficiency, $\eta_e$, is defined as the fraction of the kinetic energy
flux at the shock that is channelled into the supra-thermal and relativistic electron
component, i.e.:

\begin{equation}
0.5 \eta_e \rho_1 v_s^{3} \left(1 - \frac{1}{C^2}\right) = \epsilon_e v_2
\end{equation}

\noindent
where $\epsilon_e$ and $v_2$ are the energy density of accelerated electrons and
the downstream velocity of the flow, respectively\footnote{Note however that the efficiency in many
cases is defined without the term $\left(1-\frac{1}{C^2}\right)$ \citep[e.g.][]{2003ApJ...593..599R,2007ApJ...669..729K}.}.

We assume a power-law distribution of the accelerated electrons in momentum $Q(p) \propto p^{-\delta_{\rm{inj}}}$
with two relevant values of the injection spectral index, one consistent with the synchrotron injection
spectrum measured at the northern edge of the relic ($\delta_{\rm{inj}} = 2.6$) and another one that
is consistent with the integrated spectrum of the relic, $\alpha_{\rm{int}} = 1.1$ \citep[i.e., $\alpha_{\rm{inj}} = -0.6$, resulting in $\delta_{\rm{inj}} = 2.2$;][]{2012A&A...546A.124V}.
The required efficiency $\eta_e$ decreases as the strength of the magnetic field 
in the relic region (downstream region) is increased, as a progressively larger fraction of the shock
energy flux is radiated into the radio band via synchrotron emission; large 
values of the magnetic fields, $B \gtrsim10 \mu$G, are however excluded under the assumption that 
thermal pressure is larger than the magnetic field pressure downstream.

From Figure~\ref{fig:etarelic} we come to the extreme requirement that more than several percent of the kinetic energy
flux crossing the shock should be converted into acceleration of relativistic electrons \citep[e.g.,][]{2012ApJ...756...97K}. This argues against standard DSA and provides support for re-acceleration for which this requirement does not apply.

\subsection{Shock re-acceleration}
\label{sec:re-acceleration}
In the re-acceleration scenario, the post-shock radio spectral slope is also given by Eq.~\ref{eq:inj-mach}, unless the initial fossil spectrum has a flatter spectrum than what would be produced by Eq.~\ref{eq:inj-mach}. Therefore, we could explain the Toothbrush spectrum if we start with a fossil population with slope of $-0.8$ (i.e., $\delta=2.6$). This slope is then preserved during re-acceleration because the shock is too weak to change the slope of the particle distribution. We note that a spectral index of $-0.8$ is very typical for radio galaxies. At low enough frequencies, these particles are not yet affected by spectral ageing (and resulting change in slope). The peculiar brightness distribution of the Toothbrush relic would then reflect the distribution of fossil plasma electrons in this region. While in this case the slope of the spectrum is preserved, the normalization will increase by about a factor of about

\begin{equation}
\frac{3C}{C+2 -\delta\left(C-1\right)} 
\label{eq:norm}
\end{equation}
at the shock \citep{2005ApJ...627..733M}.

We can estimate this number to check how  bright this fossil plasma would be in comparison with the relic.
For $\delta=2.6$ (a spectral index of  $-0.8$) and a density compression factor of $C=1.26$, the normalization should increase the radio surface brightness by about a factor of 1.5 at the shock. A higher compression factor of $C=1.7$ (i.e., $\mathcal{M}=1.5$) would result in an increase in surface brightness at the shock of about $2.7$. 

Besides re-acceleration, there should also be a boost of the fossil electron spectrum due to the magnetic field compression. This will  increase the normalization by a factor of about

\begin{equation}
C^{\left(\delta+2\right)/3} \mbox{ .}
\label{eq:norm2}
\end{equation}

With $\delta=2.6$ and  $C=1.7$, this gives a boost in the surface brightness of a factor of $2.3$. It is not clear though whether this boost can be added to the one given by Eq.~\ref{eq:norm} \citep{2005ApJ...627..733M}. However, in both cases the boost is rather small. Therefore, if there is any pre-shock fossil plasma, it should only be a factor of $\lesssim 5$ times fainter in comparison with the relic.  There is a hint of some faint emission just to the north of B1, but this emission has a surface brightness  that is a factor of $\sim 10^2$ fainter than the northern part of relic B1. In addition, for the western part of the relic (B2 and B3) we do not detect pre-shock fossil plasma, whilst we have sufficient sensitivity to do so 
if it were present. This apparent lack of pre-shock fossil plasma can be explained if we assume that the fossil plasma was confined into an AGN lobe or radio tail that was stripped, accelerated, and transported downstream of the shock. In that case, the relic directly traces the remnants of a radio galaxy whose plasma has been re-accelerated. This scenario has also been invoked for the relic in the {Bullet Cluster} \citep{2015MNRAS.449.1486S} and {PLCKG287.0+32.9} \citep{2014ApJ...785....1B}. The morphology of the relic in the Bullet Cluster, and its location with respect to the X-ray emission from the ICM, is very similar to the Toothbrush cluster. However, no discrepancy between the radio and X-ray-derived Mach numbers was found by  \cite{2015MNRAS.449.1486S}.

A candidate that could have supplied this fossil plasma is the AGN located at RA 06$^{\rm{h}}$03$^{\rm{m}}$07$^{\rm{s}}$.5 Dec +42\degr16\arcmin21\arcsec.4, at the southwestern end of B1 (Figure~\ref{fig:spixhbagmrt}). This inverted spectrum radio source ($S_{\rm{610~MHz}} = 2.0\pm0.2$~mJy) is associated with a cluster member \citep[$z=0.210$;][]{dawson}. The source is unresolved at the $\approx 4\arcsec$ resolution of the GMRT 610~MHz observations \citep{2012A&A...546A.124V}, but it is located within the overall emission associated with B1.  While the source is rather faint compared to the relic, it may have been brighter in the past. However, given the spatial extent of the relic the proximity could also be circumstantial.

A problem with the above scenario of an AGN tail that was stripped and re-accelerated, is that one would expect to observe a spectral index steepening along the relic due the energy losses along the tail. Re-acceleration at a weak shock would preserve this spectral index trend. The spectral index along the northern edge of relic~B does not show clear evidence of spectral steepening from west to east (Figure~\ref{fig:spixHRwedge}). However, it should be noted that the above sketched re-acceleration scenario assumed a power-law fossil electron distribution. If the fossil electrons originated from a radio galaxy, the assumption of a power-law distribution is not valid, as the distribution is expected to have a cutoff due to energy losses. In this case, a weak shock can change the observed spectral index \citep[e.g.,][]{2001A&A...366...26E}.  Simulations will be needed to determine if this can explain the observed spectral index for the Toothbrush relic \citep[e.g.,][]{2015ApJ...809..186K}.

\subsection{A mixture of Mach numbers}
Shocks tracing radio relics should consist of a mixture of Mach numbers along the line of sight, as hydrodynamical simulations indicate \citep[e.g.,][]{2013ApJ...765...21S,2012MNRAS.421.1868V,2011ApJ...735...96S}. If  the acceleration efficiency is a non-linear function of the Mach number \citep[e.g.,][]{2007MNRAS.375...77H}, the synchrotron-weighted Mach number would naturally be higher than the X-ray weighted Mach number. In addition, if the X-ray Mach number is measured via the density jump, then the projection of different shock surfaces and shock substructures would smear out the density jump, increasing the discrepancy between the radio- and the X-ray-derived Mach numbers. 

Recent simulations of particle acceleration at cluster shocks by \cite{2015arXiv150403102H} provide some support for this explanation. The Mach numbers derived from mock radio and X-ray maps were found to differ significantly. The X-ray observations were found to  underestimate the  volume weighed shock Mach number. The derived mock radio Mach numbers gave a fairer representation of the volume averaged Mach number of shocks associated with radio relics. Therefore, (part of) the difference between the radio- and X-ray-derived Mach numbers in the Toothbrush cluster could also be explained by a mixture of Mach numbers along the line of sight. It should be noted though that these simulations did not take into account the fact that pre-existing CR electrons could be re-accelerated at weak shocks, which could influence the results.  One would also expect spectral index variations along the length of relics. For the Toothbrush relic, we do observe some spectral index variations along the northern edge of the relic (Figure~\ref{fig:spixHRwedge}). It is not clear though if these variations are consistent with these simulations, because \cite{2015arXiv150403102H} only reported on the integrated spectral indices for radio relics. Future deep X-ray and radio observations are critical to determine whether discrepancies between radio- and X-ray-derived Mach numbers of shocks tracing relics are common.

\subsection{Radio halo spectral index variations}
 \label{sec:spixhalo}
 To study the spectral index across the radio halo we extracted spectra in several regions, where the halo is well detected
by both LOFAR and the VLA, as indicated in Figure~\ref{fig:spixLRwedge}. We avoided the extreme southern part of the radio halo due to the presence of several compact sources.  The regions where the spectra were extracted correspond to boxes with physical sizes of about 110~kpc on a side.

\begin{figure}[t]
\begin{center}
\includegraphics[angle =180, trim =0cm 0cm 0cm 0cm,width=0.47\textwidth]{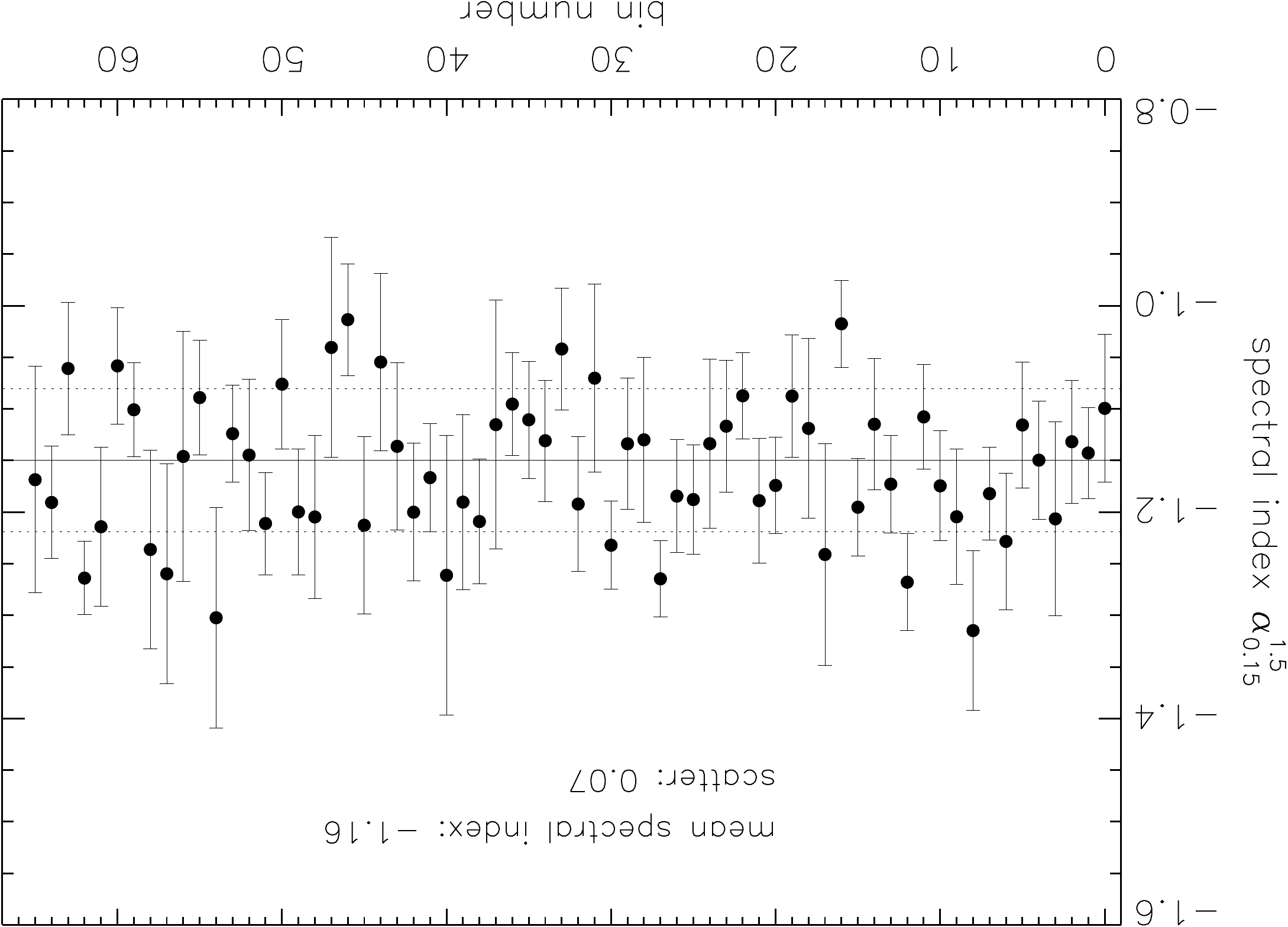}
\end{center}
\caption{Spectral index distribution across the radio  halo between 0.15 and 1.5~GHz in regions measuring 31~arcsec$^2$ (see also Figure~\ref{fig:spixhbajvla}). The regions where the spectral indices are extracted, are indicated in the left panel of Figure~\ref{fig:spixLRwedge} in blue. The solid horizontal  line indicates the average spectral index. The dotted lines represent the {measured} scatter of the spectral indices. Systematic uncertainties in the flux-scale were not included in the error bars as they do not affect the measured scatter. The region (bin) numbering was randomized.}
\label{fig:spixhalo}
\end{figure}

The spectral indices for the remaining 66 regions are displayed in Figure~\ref{fig:spixhalo}. From the 66 spectra we find that the spectral index across the radio halo over a 0.8~Mpc$^2$ region stays rather constant. Following \cite{2013ApJ...777..141C}, to properly take the measurement uncertainties into account, we compute a raw scatter from the mean spectral index of 0.07. This implies that the intrinsic spectral index scatter is $\leq 0.04$, given that {measurement} errors from map noise alone would result in a spread of 0.06 ($\sqrt{0.06^2 + 0.04^2} \approx 0.07$). The largest intrinsic variations are $\Delta \alpha \sim0.1$.
 This statement applies to the part of the radio halo that is detected at high SNR at both 1.5 and 0.15~GHz, but we note that this already covers regions that in ``deep'' old-VLA or GMRT observations would have {been} to be close to the detection limit.

The uniform spectral index is quite remarkable. In the case of \object{Abell~2256},  inhomogeneous turbulence was invoked to explain the low frequency spectral steepening \citep{2012A&A...543A..43V}. {In addition, \cite{2014A&A...561A..52V,2007A&A...467..943O,2004A&A...423..111F} find evidence for spectral variations across radio halos that are larger than for the Toothbrush. Although, it should be mentioned that the spectral index maps presented there have considerably larger spectral index uncertainties and less dense uv-coverage.} 

For the Toothbrush cluster the turbulence seems to be homogeneous. 
In the context of turbulent re-acceleration models, it would imply that the
combination of turbulent energy flux and turbulent damping on
relativistic electrons is essentially constant when averaged on 110~kpc
cells and integrated along the line of sight. It also implies that the
turbulent energy does not change significantly across the overall region
0.8~Mpc$^2$ region, that is the region just crossed by the two DM cores.
In principle these constraints are very important for the physics of
turbulence and acceleration in the ICM, however detailed  simulations
including physics of electron acceleration and turbulence are necessary to
draw meaningful conclusions.

A region with a somewhat steeper spectrum, $\alpha \simeq -1.4$ to $-1.5$, is located at the eastern part of the halo, toward relic~E (Figure~\ref{fig:spixLRwedge}). This steep-spectrum region may indicate a decline of the magnetic field or a significant steepening of the particle spectrum \citep[][]{2001MNRAS.320..365B}. However, it could also be that this steepening is due to electron energy losses downstream of relic~E. 

The spectral index variations we find for the halo are much smaller than those reported by \cite{2007A&A...467..943O} for \object{Abell~2744}. For \object{Abell~2744}, the spectral index values varied between $-0.7\pm0.1$ and $-1.5\pm0.2$. For Abell~2744, \cite{2007A&A...467..943O} also reported a  correlation of spectral index with the ICM temperature,  where the regions with the flattest spectra corresponded to regions with the highest temperature. We do not find any evidence for this correlation in RX~J0603.3+4214. A similar result was reported by \cite{2014A&A...561A..52V} for Abell~520. Either the physical conditions related to particle acceleration in Abell~2744 are different, or these variations are caused by the lower signal-to-noise of the Abell~2744 observations. We leave a detailed investigation of the physical implications of the uniform spectral index and temperature distribution for future work.

\subsection{Merger scenario}
One of the puzzling aspects of the Toothbrush cluster is the very long linear extension of relic B to the east. Although the complex shape of the relic could reflect the distribution of fossil plasma in the re-acceleration scenario, the merger still needs to produce a shock that re-accelerates this fossil plasma along the visible extent of the relic. Simulations of \cite{2012MNRAS.425L..76B} suggest that this extension could be caused by a merger between two main subclusters and a third lower mass system. In the {\it Chandra} image, we find a smaller substructure to the west of the cluster (Figures~\ref{fig:cartoon} and~\ref{fig:secimage}). Assuming it is at the same redshift as the main cluster, we compute an X-ray luminosity of $L_{\rm{X, 0.5-2.0 keV}} \approx 5 \times 10^{42}$~erg~s$^{-1}$. Using the $L_{\rm{X}}-M$ scaling relation from \cite{2009ApJ...692.1033V} this would correspond to a mass of $\sim 6\times 10^{13}$~M$_{\odot}$ which is similar to the mass used by \cite{2012MNRAS.425L..76B} of $3.5\times 10^{13}$~M$_{\odot}$.
The current location of the substructure is, however, very different from that suggested by the simulations of \cite{2012MNRAS.425L..76B}.  New simulations, taking the position of this subcluster into account, as well as weak lensing and spectroscopic measurements that provide estimates of the subcluster masses \citep{2015arXiv151003486J,dawson},  should shed more light on this issue.

\subsection{Radio halo, relic, shocks and turbulence connection}
The southern edge of the radio halo, C1, coincides with the location of the shock front. A similar situation is observed in the Bullet cluster \citep{2014MNRAS.440.2901S} and in Abell~520 {\citep{2005ApJ...627..733M,2010arXiv1010.3660M,2014A&A...561A..52V}, the Coma Cluster \citep{2013A&A...554A.140P,2015arXiv150901901U}, and Abell~754 \citep{2011ApJ...728...82M}.} Therefore, this seems to be a common phenomenon. In \cite{2012A&A...546A.124V}, we speculated whether the southern part of the radio halo (C1) could actually be a fainter relic, given the somewhat enhanced surface brightness. The rather uniform spectral index across the radio halo argues against such an interpretation. 

{Relic~D is located at the eastern end of the southern shock. The spectral index gradient across this small relic, with spectral steepening towards the north (Section~\ref{sec:spix}), is precisely what one would expect for a southwards traveling shock. Since the relic is quite small, a {largest linear size} (LLS) of 0.25~Mpc, and has a complex morphology \citep{2012A&A...546A.124V}, we speculate that this source also traces a remnant AGN lobe where particles are re-accelerated. We could not identify the source of this fossil plasma, but there are several cluster galaxies located in its vicinity.}

The radio halo seems to trace the region where the X-ray gas is highly disturbed (Figure~\ref{fig:chandra}), as the core of the southern subcluster flew through another cluster (that is now located north of it). Baryons in the ICM behave collisionally, and in major mergers after the core passage they are strongly stirred and partially decouple from the collisionless DM cores \citep[e.g.,][]{1997ApJS..109..307R,2006ApJ...648L.109C}. In this situation, fluid/kinetic instabilities and turbulence are generated which play a role in dissipating part of the kinetic energy of the gas into gas heating, particle acceleration, and magnetic field amplification.

An interesting feature is the connection between the northern relic and the halo.
A strong spectral steepening is measured downstream of the relic, while the spectrum
returns to being progressively flatter in the halo (Figures~\ref{fig:spixhbajvla} and~\ref{fig:spixLRwedge}).
This spectral behavior is suggestive of the interplay between different mechanisms.
The most appealing hypothesis is that the flattening is due to the re-acceleration by turbulence of
``aged'' electrons downstream of the relic \citep{2005ApJ...627..733M,2012mgm..conf..397M}.

On one hand, we can assume that the large scale turbulence that is responsible for the radio halo is generated by the shock passage downstream,
and that it takes time to decay into the smaller scales that are relevant for particle acceleration \citep{2014IJMPD..2330007B}.
Thus, spectral flattening would occur in the region where the decay time of large-scale gas motions, $\sim L/\delta V$ ($\delta V$ and $L$ being the
velocity and scales of the large eddies), equals the time spent by the passage of the shock, $\sim \rm{distance}$/$v_2$. This sets 
a condition $\delta V \sim 700$~km~s$^{-1}$ $\times L$/(100 kpc) (assuming $\mathcal{M} \sim 1.5$ and upstream temperature from the observations) that is qualitatively consistent with the requirements of the models of radio halos based on particle re-acceleration by compressive MHD turbulence \citep[see][]{2007MNRAS.378..245B}.

On the other hand, it is also possible that the weak shock associated with the northern relic is in fact not generating enough
turbulence and that turbulence is only present in the region where we observe spectral flattening. 
The absence of strong spectral steepening in the southern region of the halo, where the southern shock is detected, might favor this second hypothesis.
In this case the southern shock should still be traveling in a turbulent region, advecting turbulent plasma in the region associated with 
the passage of the two DM cores, while the northern shock should be traveling ahead of the DM core that is moving northward.

\section{Conclusions}
\label{sec:conclusions}
In this paper, we have presented results from deep LOFAR and {\it Chandra} observations of the Toothbrush cluster. 
The LOFAR data have allowed us to create high-quality spectral index maps, {combining with GMRT and VLA data}, and probe diffuse emission with unprecedented sensitivity in the 120--181~MHz range. To obtain these deep LOFAR images, we developed a new calibration scheme, which we name ``facet calibration'', that corrects for the time-varying complex station beams and direction dependent ionospheric effects. 
Below, we summarize our results:

\begin{itemize}

\item We obtained the first deep LOFAR images of diffuse cluster radio emission, reaching a noise level of 93~$\mu$Jy~beam$^{-1}$. These images  allowed the study of the diffuse emission down to 5\arcsec-scales for the first time below 200~MHz.

\item We detect both the radio halo and the relics. For {the} main northern Toothbrush relic, we observe a strong spectral index gradient. The spectral index is $-0.8\pm0.1$ at the northern edge of the relic and steepens to $\approx -2$ at the boundary of the relic and the halo.  The spectral index, along the northern edge of the Toothbrush relic, varies (over a distance of 1.9~Mpc) between $-0.7\pm0.1$ and $-1.0\pm0.1$, with the brighter parts corresponding to flatter spectra.

\item The radio halo has an integrated spectral index of $\alpha= -1.08 \pm 0.06$. The radio spectral index across the halo is remarkably uniform. For the main part of the radio halo, covering a region of 0.8~Mpc$^2$, we find that the intrinsic spectral index scatter is $\leq 0.04$.

\item With the {\it Chandra} observations we find a merger-related shock and cold front in the southern part of the cluster. The shock is located at the southern edge of the radio halo, 
and the shock/halo connection is similar to the Bullet Cluster.  The southern shock has a Mach number of $\mathcal{M} = 1.39^{+0.06}_{-0.06}$ measured via the density jump and $\mathcal{M} = 1.8^{+0.5}_{-0.3}$  via the temperature jump. We also find evidence for variations of the Mach number along the shock front, with the Mach number ranging from  $\mathcal{M} = 1.16^{+0.11}_{-0.07}$ to $\mathcal{M} = 1.57^{+0.16}_{-0.13}$.

\item We observe a change in the slope of the X-ray surface brightness profile at the location of the Toothbrush relic. This change of slope could indicate the presence of a very weak, $\mathcal{M}\sim1.2$, shock. The absence of a stronger shock  seems to be at odds with the radio-derived Mach number of $\mathcal{M} = 2.8^{+0.5}_{-0.3}$ predicted from DSA. Thus, the observationally-derived radio and X-ray shock Mach numbers differ significantly for this radio relic. In addition, for DSA to explain the observed radio flux density of the relic, an unrealistic fraction of the shock energy is required to produce the non-thermal electron population.

\item The apparent conflict between the X-ray derived Mach number and the one from the radio spectral index can be reconciled with a scenario where a weak shock re-accelerates a population of fossil electrons with an energy slope of $\delta \simeq 2.4$. The complex brightness distribution of the relic suggests that the source of these fossil electrons was likely an old radio galaxy. Alternatively, the discrepancy between the radio- and X-ray-derived Mach number might be explained if the relic traces multiple shock surfaces along the line of sight, coupled with the non-linear scaling of radio brightness with Mach number.

\item In the {\it Chandra} image, we detected a third, smaller, substructure. However, it is not clear if this substructure is responsible for the peculiar linear eastern extension of the main Toothbrush relic. New simulations of the merger event are required to shed more light on this issue.

\end{itemize}

\acknowledgments
{\it Acknowledgments:}
We would like to thank the anonymous referee for useful comments. 
R.J.W. was supported by NASA through the Einstein Postdoctoral
grant number PF2-130104 awarded by the Chandra X-ray Center, which is
operated by the Smithsonian Astrophysical Observatory for NASA under
contract NAS8-03060. Support for this work was provided by the National Aeronautics and Space Administration through Chandra Award Number GO3-14138X issued by the Chandra X-ray Observatory Center, which is operated by the Smithsonian Astrophysical Observatory for and on behalf of the National Aeronautics Space Administration under contract NAS8-03060. G.A.O. acknowledges support by NASA through a Hubble Fellowship grant HST-HF2-51345.001-A awarded by the Space Telescope Science Institute, which is operated by the Association of Universities for Research in Astronomy, Incorporated, under NASA contract NAS5- 26555. G.B. acknowledges support from the Alexander von Humboldt Foundation. G.B. and R.C. acknowledge support from PRIN-INAF 2014. W.R.F., C.J., and F.A-S. acknowledge support from the Smithsonian Institution. F.A.-S. acknowledges support from {Chandra} grant G03-14131X. C.F. acknowledges support by the Agence Nationale pour la Recherche, MAGELLAN project, ANR-14-CE23-0004-01. Partial support for L.R. is provided by NSF Grant AST-1211595 to the University of Minnesota. Part of this work performed under the auspices of the U.S. DOE by LLNL under Contract DE-AC52-07NA27344. LOFAR, the Low Frequency Array designed and constructed by ASTRON, has facilities in several countries, that are owned by various parties (each with their own funding sources), and that are collectively operated by the International LOFAR Telescope (ILT) foundation under a joint scientific policy. The National Radio Astronomy Observatory is a facility of the National Science Foundation operated under cooperative agreement by Associated Universities, Inc. We thank the staff of the GMRT that made these observations possible. GMRT is run by the National Centre for Radio Astrophysics of the Tata Institute of Fundamental Research. The Open University is incorporated by Royal Charter (RC 000391), an exempt charity in England \& Wales and a charity registered in Scotland (SC 038302). The Open University is authorized and regulated by the Financial Conduct Authority.

{\it Facilities:} \facility{GMRT}, \facility{CXO}, \facility{VLA}, \facility{LOFAR} 


\appendix
\section{Temperature uncertainty map and integrated flux densities}
\begin{figure*}[h!]
\begin{center}
\includegraphics[angle =180, trim =0cm 0cm 0cm 0cm,width=0.75\textwidth]{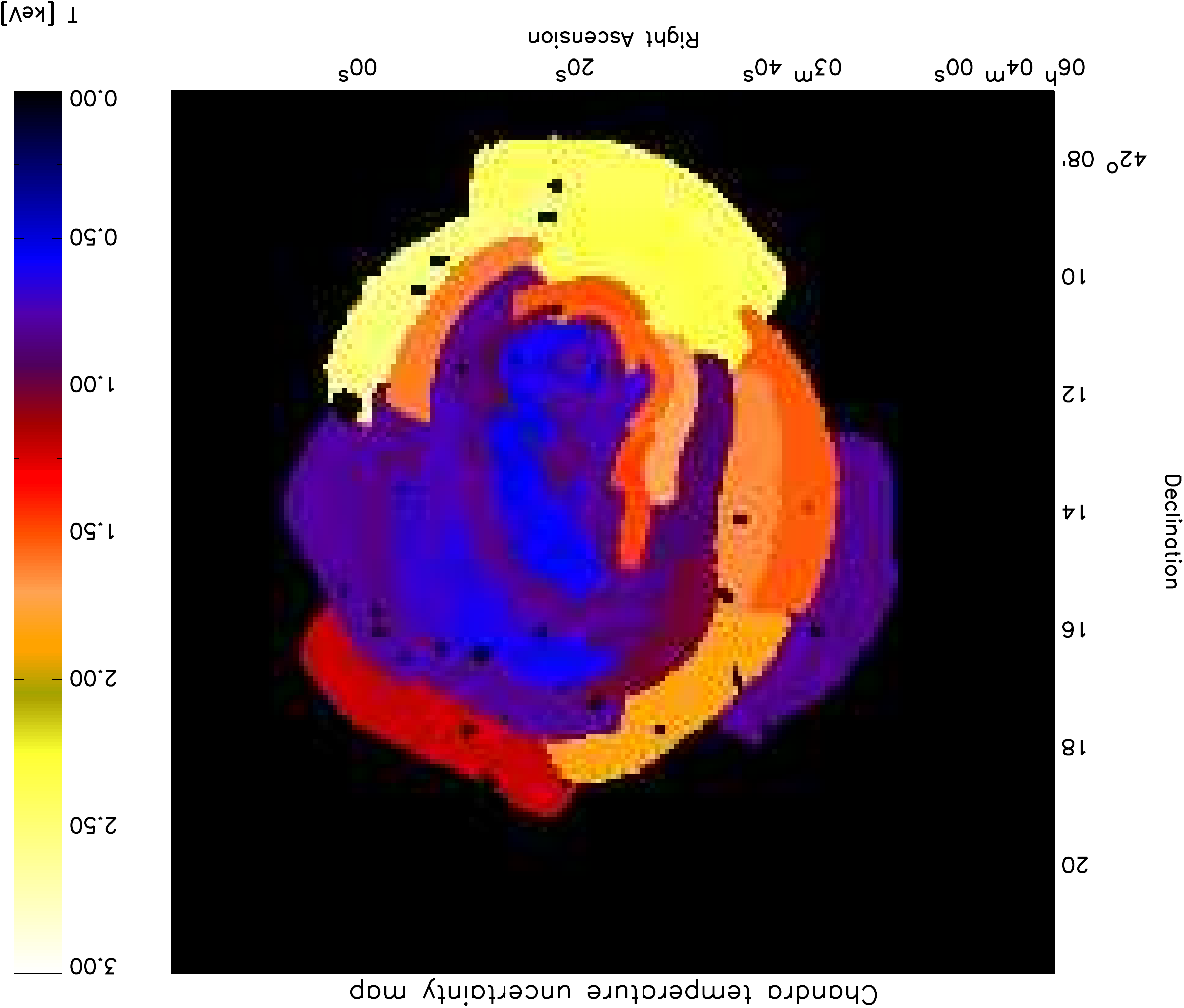} 
\end{center}
\caption{Temperature uncertainty map  corresponding to Figure~\ref{fig:tmap}. The average of the upper and lower $1\sigma$ temperature errors is shown.}
\label{fig:tmaperrs}
\end{figure*}

\begin{figure}[h!]
\begin{center}
\includegraphics[angle =0, trim =0cm 0cm 0cm 0cm,width=0.45\textwidth]{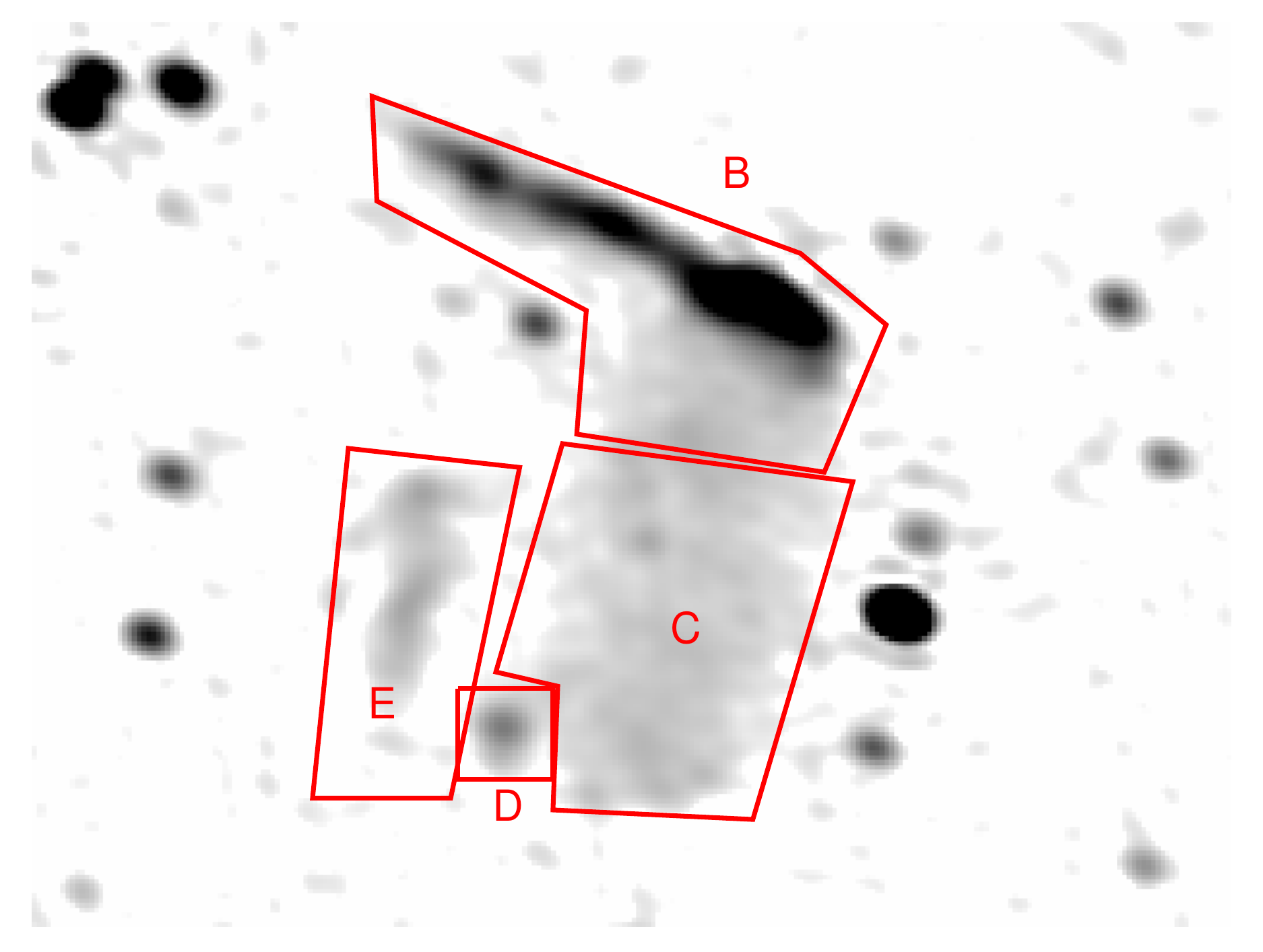}
\end{center}
\caption{VLA 1--2~GHz images depicting the regions where the integrated flux densities were measured. The image has a resolution of $31\arcsec \times 24\arcsec$.}
\label{fig:intfluxes}
\end{figure}

Figure~\ref{fig:tmaperrs} displays $1\sigma$ uncertainties in the temperature map shown in Figure~\ref{fig:tmap}. The regions where we measured the integrated flux densities at 0.15 and 1.5~GHz are depicted in Figure~\ref{fig:intfluxes}. The integrated flux densities are reported in Table~\ref{tab:rx42diffuse}.

\bibliography{ref_filaments.bib}

\end{document}